\documentclass[]{aa}

\usepackage{xcolor}  
\usepackage{natbib}  
\bibpunct{(}{)}{;}{a}{}{,} 

\usepackage[colorlinks = true,
            linkcolor = magenta,
            urlcolor  = magenta,
            citecolor = blue,
            anchorcolor = blue]{hyperref} 

\usepackage{graphicx}
\usepackage{txfonts}
\let\ACMmaketitle=\maketitle
\renewcommand{\maketitle}{\begingroup\let\footnote=\thanks \ACMmaketitle\endgroup} 

\usepackage{caption}     
\usepackage{subcaption}  

\begin{document} 
   \title{Signs of magnetic star-planet interactions in HD~118203}
   \subtitle{TESS detects stellar variability that matches the orbital period of a close-in eccentric Jupiter-sized companion}

    \author{A.~Castro-Gonz\'{a}lez\inst{ \ref{CAB_villafranca}}
    \and
    J.~Lillo-Box\inst{\ref{CAB_villafranca}}
    \and A.~C.~M.~Correia\inst{\ref{cfisuc-coimbra},\ref{imcce-paris}}
    \and
    N. C. Santos\inst{\ref{CAUP},\ref{dep_astro_porto}}
    \and 
    D. Barrado\inst{ \ref{CAB_villafranca}} 
    \and \\
    M. Morales-Calderón\inst{ \ref{CAB_villafranca}}
    \and
    E.~L.~Shkolnik\inst{\ref{asu}}
    }
    
    \institute{Centro de Astrobiolog\'{i}a, CSIC-INTA, ESAC campus, 28692 Villanueva de la Ca\~{n}ada, Madrid, Spain\label{CAB_villafranca} \\\email{acastro@cab.inta-csic.es}
    \and 
    CFisUC, Departamento de F\'isica, Universidade de Coimbra, 3004-516 Coimbra, Portugal\label{cfisuc-coimbra}
    \and 
    IMCCE, UMR8028 CNRS, Observatoire de Paris, PSL University, Sorbonne Univ., 77 av. Denfert-Rochereau, 75014 Paris, France\label{imcce-paris}
    \and
    Instituto de Astrof\'isica e Ci\^encias do Espa\c{c}o, Universidade do Porto, CAUP, Rua das Estrelas, 4150-762 Porto, Portugal\label{CAUP}
    \and
    Departamento de Fisica e Astronomia, Universidade do Porto, Rua do Campo Alegre, 4169-007 Porto, Portugal\label{dep_astro_porto}
    \and
    School of Earth and Space Exploration, Arizona State University, 660 S. Mill Ave., Tempe, Arizona 85281, USA\label{asu}
    }
    
\date{Received 23 November 2023 / Accepted 25 January 2024}

  \abstract
   {Planetary systems with close-in giant planets can experience magnetic star-planet interactions that modify the activity levels of their host stars. The induced activity is known to strongly depend on the magnetic moment of the interacting planet. Therefore, such planet-induced activity should be more readily observable in systems with close-in planets in eccentric orbits, since those planets are expected to rotate faster than in circular orbits. However, no evidence of magnetic interactions has been reported in eccentric planetary systems to date.
   }
   {We intend to unveil a possible planet-induced activity in the bright ($V$ = 8.05 $\pm$ 0.03 mag) and slightly evolved star HD~118203, which is known to host an eccentric ($e$ = 0.32 $\pm$ 0.02) and close-in ($a$ = 0.0864 $\pm$ 0.0006 au) Jupiter-sized planet.
   }
   {We characterized the planetary system by jointly modelling 56 ELODIE radial velocities and four sectors of TESS photometry. We computed the generalized Lomb-Scargle periodogram of the TESS, ELODIE, and complementary ASAS-SN data to search for planet-induced and rotation-related activity signals. We studied the possible origins of the stellar variability found, analysed its persistence and evolution, and searched for possible links with the eccentric orbital motion of HD~118203~b.
   }
   {We found evidence of an activity signal within the TESS photometry that matches the 6.1-day orbital period of its hosted planet HD~118203~b, which suggests the existence of magnetic star-planet interactions. We did not find, however, any additional activity signal that could be unambiguously interpreted as the rotation of the star, so we cannot discard stellar rotation as the actual source of the signal found. Nevertheless, both the evolved nature of the star and the significant orbital eccentricity make the synchronous stellar rotation with the planetary orbit very unlikely.
   }
   {The planetary system HD~118203 represents the best evidence that magnetic star-planet interactions can be found in eccentric planetary systems, and it opens the door to future dedicated searches in such systems that will allow us to better understand the interplay between close-in giant planets and their host stars.
   }

   \maketitle
%

\section{Introduction}

In the late 1980s,  \citet{1988ApJ...331..902C} presented the first exoplanet candidate that was later found to be a real planet \citep{2003ApJ...599.1383H}. Soon after, \citet[][]{1995Natur.378..355M} confirmed the first exoplanet orbiting a Sun-like star, 51 Peg b. Having half the mass of Jupiter and orbiting its star every 4.2 days, 51 Peg b was the prelude to the detection of tens of similar planets through the radial velocity technique \citep[e.g.][]{1997ApJ...474L.115B,1998PASP..110.1389B,2000A&A...356..599S} and the transit method \citep[e.g.][]{2000ApJ...529L..41H,2000ApJ...529L..45C}. Those giant planets orbiting close to their host stars ($\lesssim$ 0.1 au) are commonly known as hot Jupiters, and they dominated the exoplanet demographics in the early years of exoplanet exploration since they were the easiest to detect. 

The \textit{Kepler} mission \citep{2010Sci...327..977B,2014PASP..126..398H} detected thousands of small planets thanks to several statistical validation works that required minimal follow-up observations \citep[e.g.][]{2018AJ....156..277L,2018AJ....156...78L,2020MNRAS.499.5416C,2021MNRAS.508..195D,2022AJ....163..244C}. Interestingly, there was not such an explosive increase for the hot Jupiter detections because of two main reasons. First, planet occurrence studies revealed that hot Jupiters are an unusual outcome of planet formation and evolution \citep[e.g.][]{2011arXiv1109.2497M,2012ApJS..201...15H}. Second, much of the radial velocity follow-up effort was focussed on the newly discovered Neptune-sized and super-Earth-sized planets \citep[e.g.][]{2020A&A...640A..48L,2023AJ....166...49D,2023MNRAS.523.3069O,2024arXiv240106276P}, thus leaving thousands of hot Jupiter candidates detected via transit without confirmation\footnote{According to the NASA Exoplanet Archive \citep{2013PASP..125..989A}, 3176 planet candidates with radii larger than 10 $R_{\oplus}$ and orbital periods smaller than 10 days remain unconfirmed.}. 

Jupiter-sized planets, however, have played a major role in our understanding of the properties and formation of planetary systems. 
Shortly after the first discoveries, following initial suspicions \citep{1997MNRAS.285..403G}, \citet{2001A&A...373.1019S,2004A&A...415.1153S} found that the fraction of gas giants increases as the stellar host metallicity increases, which favours the traditional core-accretion planet formation theory against the alternative disk instability theory. Also, hot Jupiters provide ideal laboratories to study tidal interactions and magnetic star-planet interactions (MSPIs) that do not exist in our own Solar System. This is because hot Jupiters typically lie inside the Alfvén radius of their host stars, which marks the boundary between the stellar corona and stellar wind  ($\simeq$~10~$R_{\star}$). At such short orbital distances, the coronal plasma's Alfvén velocity is larger than the stellar wind velocity, which allows for a direct magnetic interaction with the stellar surface.

The possibility of tidal interactions and MSPIs in planetary systems hosting hot Jupiters was first suggested by \citet{2000ApJ...533L.151C}. Magnetic star-planet interactions can be manifested as periodic stellar activity that matches the orbital period of the interacting planet. In contrast, tidal interactions can induce stellar variability with half the orbital period. One of the first signs of MSPIs was observed in HD 192263, which was known to have a hot Jupiter detected via radial velocities by \cite{2000A&A...356..599S}. This planet was put into question by \citet{2002ApJ...577L.111H}, who detected periodic photometric variations with a very similar period to that of the reported planetary orbit. In a subsequent study, \citet{2003A&A...406..373S} acquired simultaneous radial velocity and photometric data and found that the photometric variations alternated stable and variable moments. In contrast, the radial velocity variations showed long-term stability in period, phase, and amplitude. This result supported the idea that the radial velocity signal had a planetary origin and that the photometric signal could be produced either by the rotation of the star or by magnetic interactions induced by the hosted planet. Only four months after, \citet{2003ApJ...597.1092S} reported the first piece of evidence of MSPIs in HD 179949, which was known to host a 3-day orbit hot Jupiter \citep{2001ApJ...551..507T}. The authors detected periodic chromospheric activity through Ca II H and K variability that matched the orbital period of the planet, as well as 7-day stellar variability corresponding to the rotation of the star. Since this first discovery, dozens of MSPIs have been reported and analysed \citep[e.g.][]{2005ApJ...622.1075S,2008ApJ...676..628S,2008A&A...482..691W,2009EM&P..105..373P,2011ApJ...741L..18P,2015ApJ...811L...2M,2019NatAs...3.1128C}. Tidal interactions, however, took longer to be confirmed. One of the first examples was the case of Kepler-91~b, in which \citet{2014A&A...568L...1L,2014A&A...562A.109L} found evidence of stellar tidal distortions within the \textit{Kepler} light curve. More recently, \citet{2022A&A...657A..52B} detected for the first time tidal distortions in an exoplanet, and \citet{2022MNRAS.513.4380I} found that tidal forces exerted by hot Jupiters can alter the rotation evolution of their host stars.

Direct detections of hot Jupiter's magnetic fields have been attempted through radio observations aiming at observing the electron cyclotron maser instability \citep{1998JGR...10320159Z,2006A&ARv..13..229T}. Recently, \citet{2022NatAs...6..141B} reported signatures of magnetization in HAT-P-11 b through a detection of a magnetotail, but previous attempts to detect radio emission from planetary magnetic fields were not successful \citep[e.g.][]{2000ApJ...545.1058B,2003ASPC..294..151F,2004IAUS..213...73F,2006pre6.conf..595W,2007MNRAS.382..455G,2008MNRAS.390..741G,2009A&A...500L..51L,2010AJ....140.1929L,2012MNRAS.423.3285V,2015aska.confE.120Z}. To date, the best approach to probe planetary magnetic fields is through indirect MSPIs. Although they have been only found in systems hosting hot Jupiters, their detection in systems with smaller planets might also be possible in the near future. It will be particularly relevant to detect MSPIs generated by small planets in the habitable zones (HZs) of their stars \citep{1993Icar..101..108K,2013ApJ...765..131K} since planetary magnetic fields are likely a necessary condition for surface habitability. Interestingly, the star-planet separations where MSPIs can appear begin to coincide with the HZs of M-dwarf and some K-dwarf stars, being the latter recently proposed as the most suitable stars to search for life beyond Earth \citep{2019ApJ...872...17R,2022A&A...667A.102L}.

In order to understand planetary magnetic fields it is important to study under what conditions they appear. In the Solar System, there is a strong correlation between the magnetic moment of a body and the ratio between its mass and its rotation period \citep[e.g.][]{2002Icar..157..507K}. Similarly, \citet{2003ApJ...597.1092S,2005ApJ...622.1075S,2008ApJ...676..628S} found that the stellar variability induced by MSPIs is also correlated with such a ratio, which suggests that MSPIs are dominated by the planetary magnetic moment as well. This observational evidence is supported by several formalisms \citep[e.g.][]{2009A&A...505..339L,2012A&A...544A..23L,2013A&A...552A.119S,2015ApJ...815..111S}. 

The rotation periods of exoplanets are difficult to measure. However, they can be confidently estimated for hot Jupiters since those planets undergo strong and well-known tidal forces that affect both their orbits and rotation velocities. In particular, their orbits tend to circularize and their rotation periods tend to synchronize with the planetary orbital period in very short time scales \citep[i.e. of a few million years;][]{1981A&A....99..126H,2018AJ....155..118W}. To date, all MSPIs have been found in circular planetary systems, which allowed us to assume that the orbital and rotation periods coincide. However, there are a few exceptions in which hot Jupiters managed to preserve an eccentric orbit because of particular conditions such as planet migration or gravitational interactions induced by an outer companion. In such eccentric systems, the planetary rotation periods are not expected to be synchronized, but rather pseudo-synchronized with the orbital periods \citep{1981A&A....99..126H,Correia_etal_2011}. This pseudo-synchronization depends on the planet's eccentricity and translates into higher rotation velocities that would generate larger magnetic planetary moments than those in circular systems. Interestingly, eccentric planetary systems also offer a unique opportunity to probe the possible imprint of the orbital geometry into the planet-induced activity signals. Unfortunately, all the attempts to detect MSPIs in eccentric systems have turned out to be unsuccessful \citep[e.g.][]{2014IAUS..299..291H,2016A&A...592A.143F}. To date, there is only one case of periodic activity \citep{2015ApJ...803...49Q} and one case of sudden activity enhancement \citep{2015ApJ...811L...2M} in eccentric systems that could be explained by MSPIs.

In this work, we aim to search for MSPIs in the bright ($V$ = 8.05 $\pm$ 0.03 mag) and slightly evolved star HD~118203, which is known to host the eccentric ($e$ = 0.32 $\pm$ 0.02) and close-in ($a$ = 0.0864 $\pm$ 0.0006 au) transiting Jupiter-sized planet HD~118203~b \citep{2006A&A...446..717D,2020AJ....159..243P}. In Sect.~\ref{sec:observations}, we describe the observations. In Sect.~\ref{sec:analysis_results}, we present our analysis of photometric and spectroscopic time series. In Sect.~\ref{sec:discussion}, we discuss the results found, and we conclude in Sect.~\ref{sec:conclusions}.


\section{Observations}
\label{sec:observations}

\subsection{TESS photometry}
\label{sec:obs_tess}

\begin{figure}
    \centering
    \includegraphics[width = 9cm]{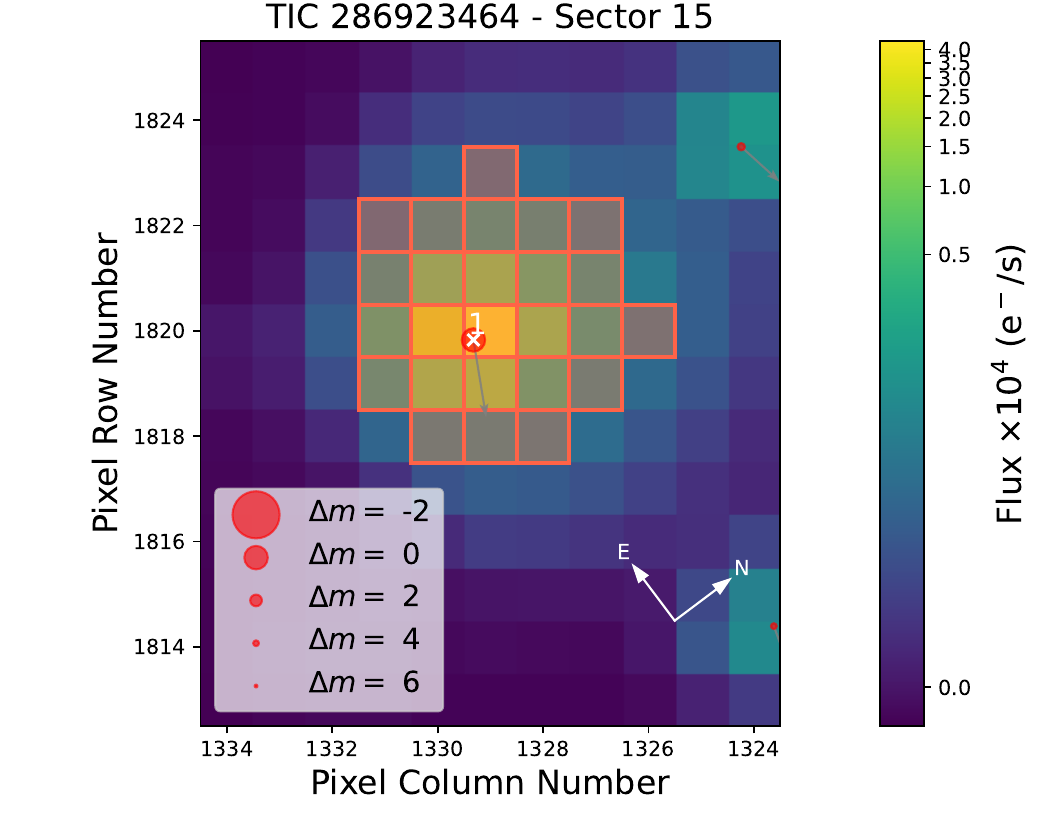}
    \caption{TESS target pixel file of HD 118203. The orange grid is the selected aperture and the red circles correspond to the nearby \textit{Gaia} DR3 sources scaled to their $G$ magnitudes.}
    \label{fig:tpfplotter}
\end{figure}

The star HD 118203 (TOI-1271, TIC 286923464) has been observed by the Transiting Exoplanet Survey Satellite \citep[TESS;][]{2015JATIS...1a4003R} in sectors 15, 16, 22, and 49 at a 2-min cadence, resulting in a 3-month duty cycle obtained throughout a 2.5-year baseline. In total, 61556 target pixel files (TPFs) were acquired, of which 9502 were discarded as having a non-zero quality flag (QF). In Table \ref{tab:TESS_summary} we summarize the details of the observations. We downloaded the presearch data conditioned simple aperture photometry \citep[PDCSAP;][]{2012PASP..124.1000S,2012PASP..124..985S,2014PASP..126..100S} from the Mikulski Archive for Space Telescopes (MAST)\footnote{\url{https://mast.stsci.edu/portal/Mashup/Clients/Mast/Portal.html}}, which was processed by the Science Processing
Operation Center (SPOC) pipeline \citep{2016SPIE.9913E..3EJ}. The SPOC module Create Optimal Aperture (COA) selected the photometric aperture that maximized the signal-to-noise of the measured flux, and then estimated the flux fraction inside the aperture that came from the target star (99.96$\%$) in order to correct for possible photometric contamination. In Figure \ref{fig:tpfplotter}, we used \texttt{tpfplotter}\footnote{\url{https://github.com/jlillo/tpfplotter}} \citep{2020A&A...635A.128A} to plot the selected aperture over a TPF of HD~118203, together with all the nearby stars detected in the \textit{Gaia} Data Release 3 \citep[DR3;][]{2022arXiv220800211G}. There are no additional sources within the aperture with a magnitude difference $\Delta$G $<$ 6 mag with HD 118203 in the \textit{Gaia} passband, which implies a negligible contamination \citep[e.g.][]{2022MNRAS.509.1075C}. In October 2019, the Transit Planet Search (TPS) algorithm of the SPOC pipeline detected a periodic flux decrease of 3.86 ppt (parts per thousand) every 6.13 days, which passed all the diagnostic tests described in \citet{Twicken:DVdiagnostics2018}. Both the orbital period and mid-transit time coincide with that of the RV-detected planet HD 118203~b, which confirms its transiting nature. The out-of-transit photometry of this target has a standard deviation of 0.46 ppt, which makes it an ideal target for searching for low-amplitude planet-induced activity signals. 

\begin{table*}[]
\caption{Summary of the TESS observations of HD 118203.}
\renewcommand{\arraystretch}{1.3}
\setlength{\tabcolsep}{16.3pt}
\begin{tabular}{ccccccc}
\hline \hline
Sector & Start date        & End date        & TPFs  & Camera & CCD & $\rm NZQF^{(1)}$ \\ \hline
15     & 15 August 2019    & 9 September 2019 & 15256 & 4      & 4   & 2493   \\
16     & 12 September 2019 & 5 October 2019   & 14678 & 4      & 3   & 2041   \\
22     & 20 February 2020  & 17 March 2020    & 17815 & 3      & 2   & 804    \\
49     & 1 March 2022      & 25 March 2022    & 13807 & 3      & 1   & 4164   \\ \hline
\end{tabular}
\tablefoot{(1) NZQF stands for the number of TESS observations with a non-zero quality flag.}
\label{tab:TESS_summary}
\end{table*}

\subsection{ELODIE spectoscopy}
\label{sec:obs_elodie}

HD 118203 was observed by the ELODIE spectrograph \citep{1996A&AS..119..373B}, which was in operation between 1993 and 2006 at the 1.93~m telescope located at the Observatoire de Haute-Provence, France. ELODIE was a fibre-fed echelle spectrograph placed in a temperature-controlled room, with a resolution power of 42000 and a wavelength range from 389.5 to 681.5 nm split into 67 spectral orders. A total of 43 spectra were acquired between May 2004 and July 2005 with a typical cadence of 1 day and a typical exposure time of 1200 s. These observations were analysed in the discovery paper \citep{2006A&A...446..717D}. In addition, 13 more spectra were acquired right after the publication of \citealp{2006A&A...446..717D} (between March and June 2006). In this work, we analyse all the 56 ELODIE spectra, which have a median signal-to-noise ratio of 43.5 at 555 nm, and are publicly available on the Data $\&$ Analysis Center for Exoplanets (DACE)\footnote{\url{https://dace.unige.ch/dashboard/}}. The spectra were reduced through the ELODIE Data Reduction Software (TACOS), which also extracted the RVs and activity indicators such as the full width at half maximum (FWHM) and the contrast of the cross-correlation function (CCF) through the cross-correlation technique \citep{1996A&AS..119..373B}. We used these indicators to search for planet-induced and rotation-related activity signals. 

\subsection{ASAS-SN photometry}
\label{sec:asas_sn}

HD 118203 has been monitored by four stations of the All-Sky Survey for Supernovae \citep[ASAS-SN;][]{2014ApJ...788...48S,2017PASP..129j4502K}. Each station consists of four Nikon telephoto lenses of 14 cm aperture, which are equipped with a 2048 $\times$ 2048 pixels CCD camera with a pixel scale of 8.0 arc seconds. The survey pipeline  performs aperture photometry through \texttt{IRAF} \citep{1986SPIE..627..733T} considering a 2-pixel radius aperture and 7-10 pixel radius annulus for the target star and for the reference stars, which are selected from the Photometric All-Sky Survey \citep[APASS,][]{2012JAVSO..40..430H,2019JAVSO..47..130H}.

We used the ASAS-SN Sky Patrol web interface\footnote{\url{https://asas-sn.osu.edu/}} to compute and download the light curves of HD 118203. This target is a high proper motion star ($\mu_{\alpha}$ = $-85.88 \pm 0.05$ $\rm mas\,yr^{-1}$, $\mu_{\delta}$ = $-78.91 \pm 0.04$ $\rm mas\,yr^{-1}$; \citealp{2022arXiv220800211G}). Hence, given that the ASAS-SN photometry is extracted in a fixed location of the detector, and the typical FWHMs are comparable to the radius of the aperture, we shifted the aperture location over time in order to minimize possible flux losses. HD 118203 is visible from the four stations from 15 November to 1 September. Hence, similarly to \citet{2023A&A...675A..52C}, we computed the photometry of HD 118203 considering the coordinates corrected for proper motion corresponding to the central time of each observing window. Finally, we discarded those epochs in which the flux is below the estimated 5$\sigma$ detection limit for the target location, as well as those data points with a deviation greater than 5$\sigma$ of a flattened version of the photometric time series, which could be caused by flares, cosmic rays, or uncorrected instrumental systematics. The median standard deviation of the final ASAS-SN photometry of HD 118203 is 15.5 ppt.


\section{Analysis and results}
\label{sec:analysis_results}

\subsection{Stellar characterization}

\begin{figure}
    \centering
    \includegraphics[width = 0.5\textwidth]{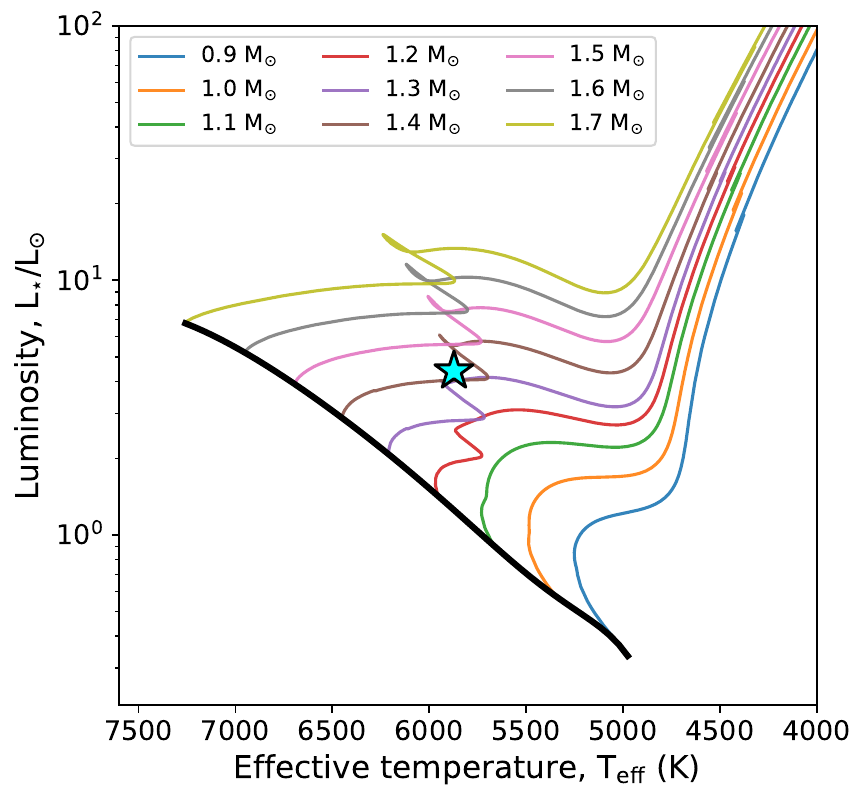}
    \caption{Location of HD~118203 in the Hertzsprung–Russell diagram. The black curve represents the main sequence, and the coloured curves are the PARSEC 2.1s tracks from \citet{2012MNRAS.427..127B,2013EPJWC..4303001B} for stars with Z = 0.04 ([M/H] $\approx$ +0.3). }
    \label{fig:HR}
\end{figure}

Given its brightness \citep[V = 8.05 $\pm$ 0.03 mag;][]{2019JAVSO..47..130H} and the early discovery of an orbiting planet \citep{2006A&A...446..717D}, HD~118203 has been extensively studied by several groups which were mainly focussed on determining elemental abundances of planet-hosting stars \citep[e.g.][]{2011ApJ...738...97B,2011A&A...530A..54G,2013A&A...556A.150S,2013A&A...554A..84M,2015A&A...576A..94S,2015A&A...576A..69D,2016A&A...588A..98M,2017AJ....153...21L,2018A&A...612A..93M}. While a comprehensive stellar characterization is out of the scope of this work, we need to adopt a stellar mass, radius, and effective temperature to determine the orbital and physical properties of HD~118203~b. We chose the spectroscopic values from the catalogue of Stars With ExoplanETs \citep[SWEET-Cat;][]{2021A&A...656A..53S}. In Table \ref{tab:stellar_params} we summarize the main properties of the star. The effective temperature and log $g$ indicate that the star is slightly evolved. In Fig.~\ref{fig:HR} we place  HD~118203 within the Hertzsprung–Russell diagram, which illustrates the star ascending the subgiant branch.

\begin{table}[]
\caption{Stellar properties of HD 118203.}
\label{tab:stellar_params}
\renewcommand{\arraystretch}{1.2}
\setlength{\tabcolsep}{5pt}
\begin{tabular}{llc}
\hline \hline
Parameter & Value & Reference \\ \hline
\multicolumn{3}{l}{Identifiers} \\ \hline
HD & 118203 & [1]  \\
TOI & 1271 & [2] \\
TIC & 286923464 & [3] \\
2MASS & J13340254+5343426 & [4] \\
Gaia DR3 & 1560420854826284928 & [5] \\ \hline
\multicolumn{3}{l}{Coordinates, parallax and kinematics} \\ \hline
RA, DEC & 13:34:02.39, 53:43:41.48 & [5] \\
$\rm \mu_{\alpha}$ ($\rm mas\,yr^{-1}$) & -85.877 $\pm$ 0.052 & [5] \\
$\rm \mu_{\delta}$ ($\rm mas\,yr^{-1}$) & -78.913 $\pm$ 0.038 & [5] \\
Parallax (mas) & 10.864 $\pm$ 0.018 & [5] \\
Distance (pc) & 92.04 $\pm$ 0.15 & [5] \\
RV ($\rm km\,s^{-1}$) & -29.37 $\pm$ 0.13 & [5] \\ \hline
\multicolumn{3}{l}{Atmospheric parameters} \\ \hline
$T_{\rm eff}$ (K) & 5872 $\pm$ 20 & [6] \\
log $g$ (dex) & 4.05 $\pm$ 0.04 & [6] \\
$\rm [Fe/H]$ (dex) & 0.27 $\pm$ 0.02 & [6] \\
\hline
Physical parameters &  &  \\ \hline
$\rm R_{\star}$ ($\rm R_{\odot}$) & 1.993 $\pm$ 0.065 & [6] \\
$\rm M_{\star}$  ($\rm M_{\odot}$) & 1.353 $\pm$ 0.006 & [6] \\
L ($\rm log_{10}$ $\rm L_{\odot}$) & $0.6458\pm 0.0018$ & [5] \\ \hline
Magnitudes &  &  \\ \hline
TESS (mag) & 7.4556 $\pm$ 0.0060 & [3] \\
G (mag) & 7.89255 $\pm$ 0.00031 & [5] \\
B (mag) & 8.746 $\pm$ 0.025 & [7] \\
V (mag) & 8.050  $\pm$ 0.030 & [7] \\
J (mag) & 6.861 $\pm$ 0.021 & [4] \\
H (mag) & 6.608 $\pm$ 0.038 & [4] \\
$\rm K_{s}$ (mag) & 6.543 $\pm$ 0.023 & [4] \\ 
\hline
\end{tabular}
\tablefoot{[1] \citet{1918AnHar..91....1C}; [2] \citet{2021ApJS..254...39G}; [3] \citet{2019AJ....158..138S}; [4] \citet{skrutskie2006}; [5] \citet{2022arXiv220800211G}; [6] \citet{2021A&A...656A..53S}; [7] \citet{2019JAVSO..47..130H}.}

\end{table}

\subsection{Joint radial velocity and transit model}
\label{sec:joint_fit}

\begin{figure*}
    \centering
     \includegraphics[width = \textwidth]{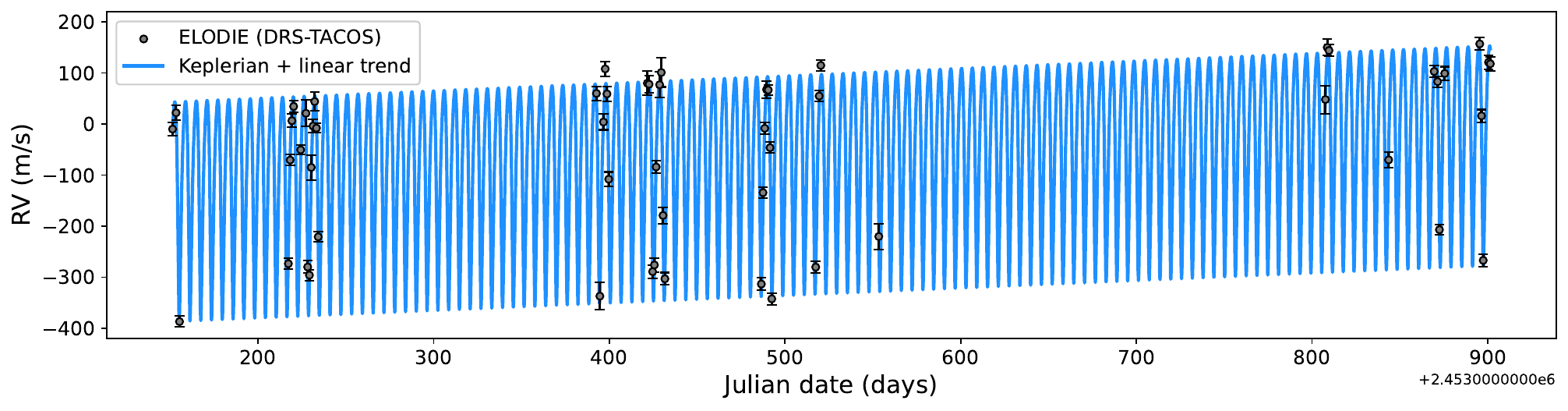}
     \caption{ELODIE RVs of HD 118203 with the median posterior model (Keplerian + linear trend) superimposed.}
     \label{fig:rv_complete}
\end{figure*}

\begin{figure*}
    \centering
     \includegraphics[width = \textwidth]{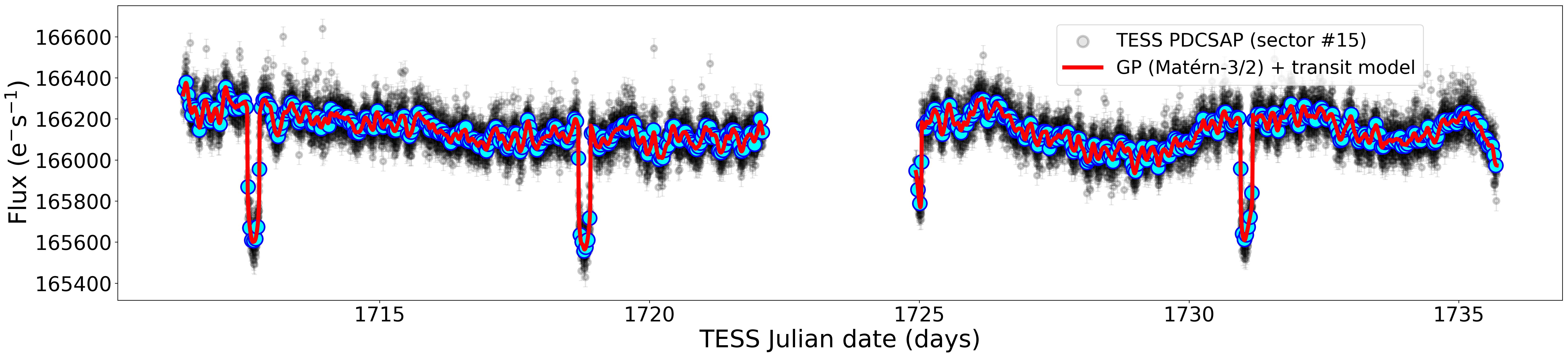}
    \includegraphics[width = \textwidth]{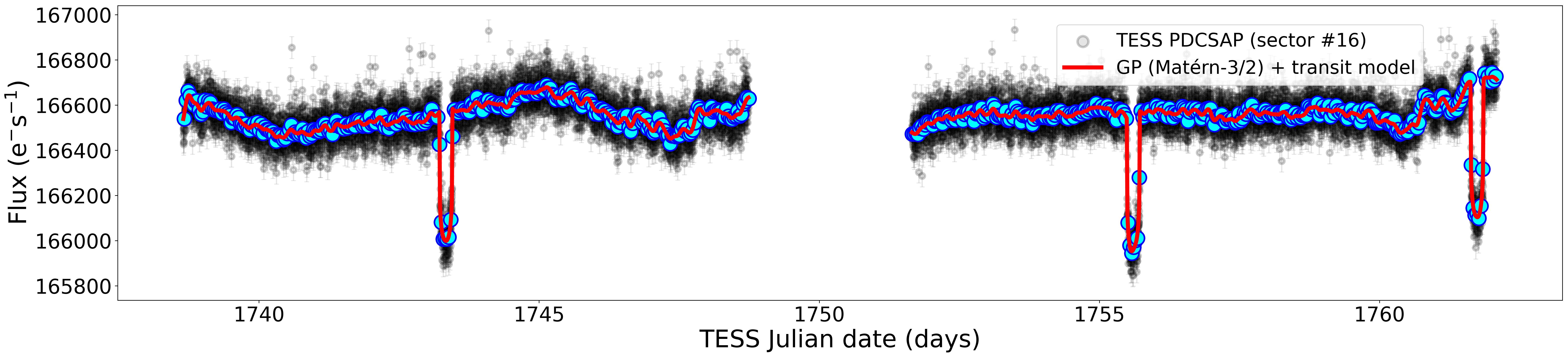}
    \includegraphics[width = \textwidth]{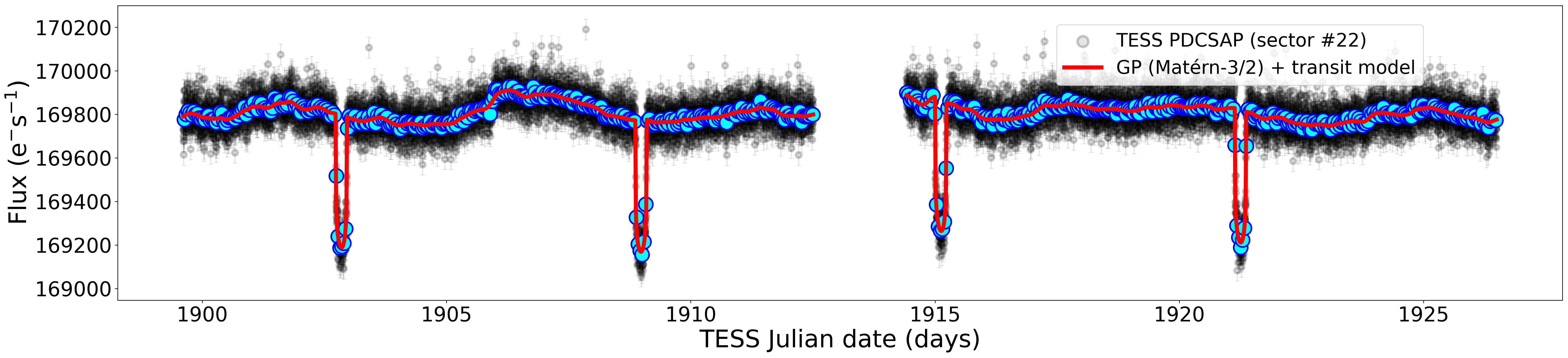}
    \includegraphics[width = \textwidth]{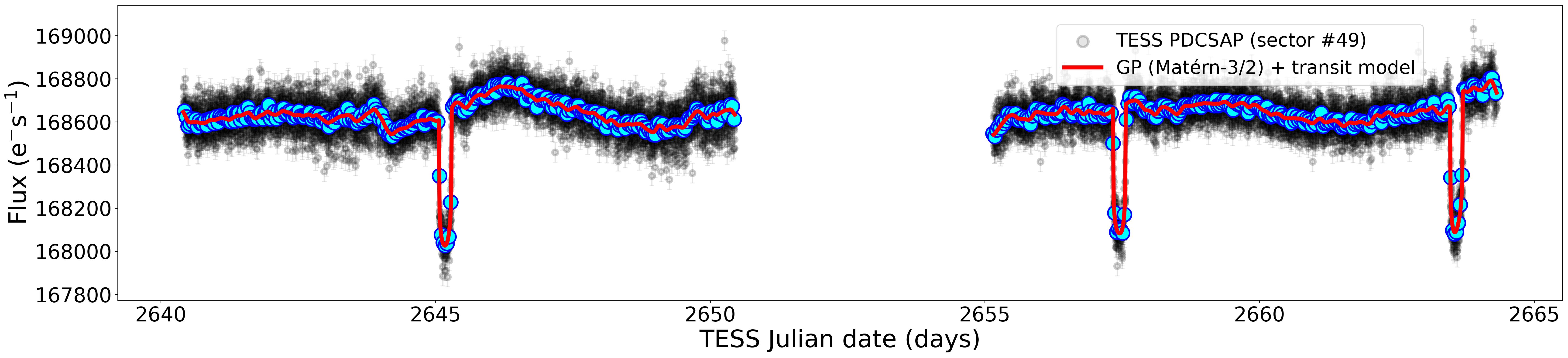}
     \caption{TESS photometry of HD 118203 with the median posterior model (transit + GP) superimposed. The grey data points correspond to the SPOC 2-minute PDCSAP photometry, and the blue data points correspond to 50-minute binned data.}
     \label{fig:TESS_complete}
\end{figure*}

\begin{figure*}
    \centering
    \includegraphics[width = 0.45\textwidth]{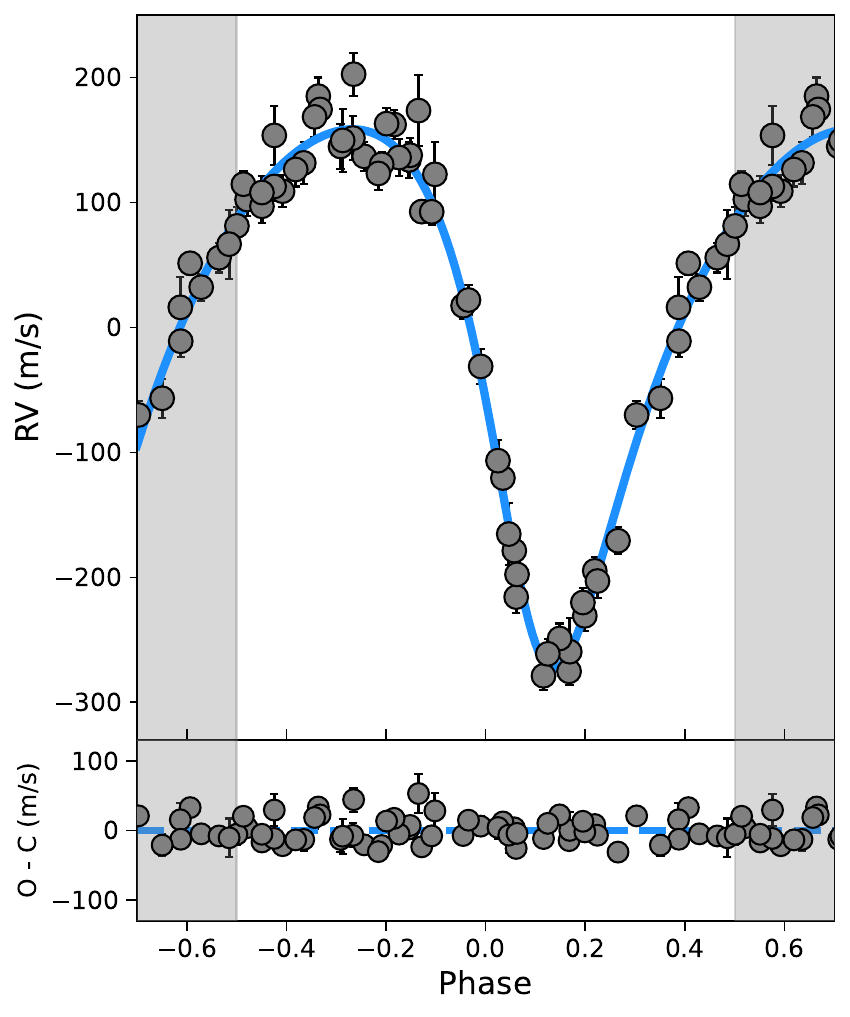}
    \includegraphics[width = 0.463\textwidth]{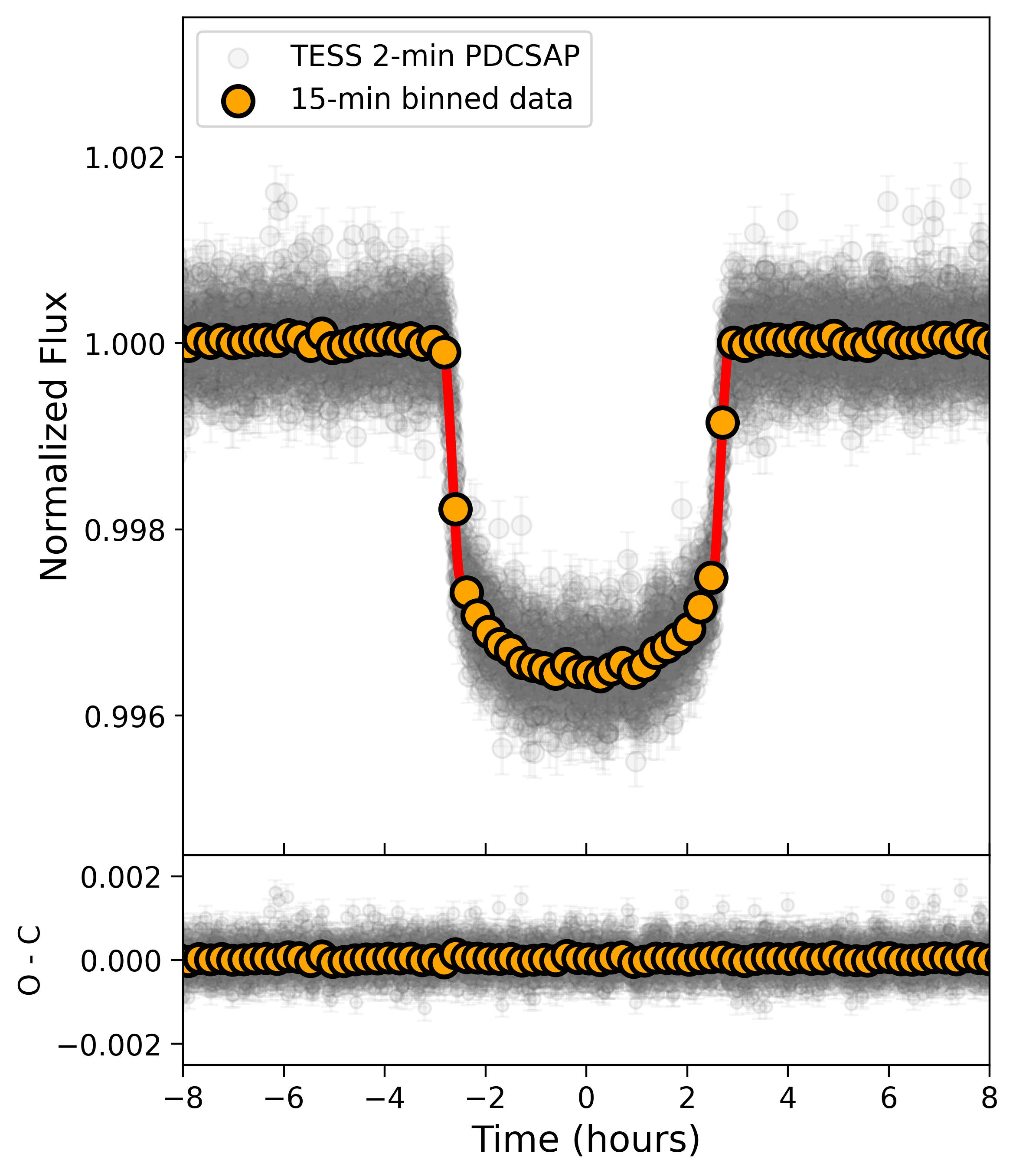}
    \caption{ELODIE RVs (left) and TESS photometry (right) folded  to the orbital period of HD~118203~b. ELODIE data were subtracted from the linear trend, and TESS data were subtracted from the GP component. The solid lines indicate the median posterior models.}
    \label{fig:folded_rvs_and_phot}
\end{figure*}

We inferred the planetary and orbital parameters of HD~118203~b by modelling jointly the TESS photometry and ELODIE RVs described in Sects. \ref{sec:obs_tess} and \ref{sec:obs_elodie}, respectively. The RVs are not significantly influenced by the stellar activity, but the TESS photometry shows a clear modulation in all the observed sectors (see Sects. \ref{sec:var_tess} and \ref{sec:variability_ELODIE} for a detailed analysis of the variability of both datasets). Hence, we modelled the ELODIE RVs with a simple Keplerian, and the TESS photometry with a model composed of a transit model and a Gaussian Process \citep[GP;][]{2006gpml.book.....R,2012RSPTA.37110550R}. Modelling planetary transits together with correlated photometric noise (either of stellar or instrumental origin) has been shown to be the most optimal procedure to preserve the transit shapes \citep[e.g][]{2021A&A...645A..41L,2023A&A...669A.109L,2023A&A...679A..33D} and to properly propagate the uncertainties of the model parameters \citep{2021A&A...649A..26L}. In the following, we describe each model as well as the fitting procedure.

We implemented the Keplerian model based on the \texttt{radvel} package \citep{2018PASP..130d4504F}. We considered the parametrization $\left\lbrace P_{\rm orb}, T_{0}, K, \sqrt{e} \cos(w), \sqrt{e} \sin(w)\right\rbrace$ recommended by \citet{2013PASP..125...83E}, where $P_{\rm orb}$ is the orbital period of the planet, $T_{0}$ the time of inferior conjuntion, $K$ the semiamplitude, $e$ the orbital eccentricity, and $w$ the planetary argument of the periastron. We included a linear trend component in order to account for a possible long-period drift. This component is described by the systemic radial velocity of the star ($v_{\rm sys}$) and a slope ($\gamma$). Finally, in order to model the white noise not taken into account in our model, we included a jitter term ($\sigma_{\rm ELODIE}$) that we added quadratically to the uncertainties of the RV measurements. We implemented the \citet{2002ApJ...580L.171M} quadratic limb darkened transit model through \texttt{batman} \citep{2015PASP..127.1161K}, which is described by the $P_{\rm orb}$, $T_{0}$, $e$, $w$, the orbital inclination ($i$), the quadratic limb darkening (LD) coefficients $u_{1}$ and  $u_{2}$, the planet-to-star radius ratio ($R_{\rm p}/R_{\star}$), and the semimajor axis scaled to the stellar radius, which we parametrized through $P_{\rm orb}$ and the stellar mass ($M_{\rm \star}$) and radius ($R_{\rm \star}$) following the Kepler's Third Law. We implemented a GP defined by an approximate Matérn-3/2 kernel through \texttt{celerite} \citep{2017AJ....154..220F,2018RNAAS...2...31F}, which has been extensively used to model TESS photometry given its simplicity and flexibility. This GP kernel can be written as:

\begin{equation}
    K_{3/2} = \eta_{\sigma}^{2} \left[ \left(1 + \frac{1}{\epsilon} \right) e^{-(1-\epsilon) \sqrt{3} \tau / \eta_{\rho}} \cdot \left(1 - \frac{1}{\epsilon} \right) e^{-(1+\epsilon) \sqrt{3} \tau / \eta_{\rho}} \right],
\end{equation}

\noindent where the hyperparameters $\eta_{\sigma}$ and $\eta_{\rho}$ are the characteristic amplitude and timescale of the correlated variations, respectively, and $\epsilon$ controls the approximation to the exact Matérn-3/2 kernel. Since both the amplitude and timescale of the TESS variability can vary from one sector to another \citep[e.g.][]{2023A&A...677A.182M},  we fitted those parameters independently; that is, we fitted $\eta_{\rm \sigma_{i}}$ and $\eta_{\rm \rho_{i}}$, where \textit{i} indicates the sector. We also included an offset ($F_{\rm 0,i}$) and a jitter term ($\sigma_{\rm TESS,i}$) for each sector in order to model the white noise not taken into account in our model.

We sampled the posterior probability density function of the different parameters involved in our global model by using a Markov Chain Monte Carlo (MCMC) affine-invariant ensemble sampler \citep{2010CAMCS...5...65G} as implemented in \texttt{emcee} \citep{2013PASP..125..306F}. We used 240 walkers (eight times as many as the number of parameters), and performed two consecutive runs. The first run (or burn-in) consisted of 200\,000 iterations. After this run, we reset the sampler and initialized the second run (or production) with 100\,000 iterations while considering the initial values from the last iteration of the burn-in phase. We finally estimated the autocorrelation time for each parameter and checked that it is at least 50 times smaller than the chain length, which indicates that we collected thousands of independent samples after discarding the burn-in phase. In Table~\ref{tab:bestfit}, we include the priors used in the MCMC run, together with the median and 1$\sigma$ (68.3$\%$ credible intervals) of the posterior distributions of the fitted parameters. In Figs.~\ref{fig:rv_complete} and \ref{fig:TESS_complete}, we show the complete ELODIE and TESS datasets together with the median posterior global model. In Fig. \ref{fig:folded_rvs_and_phot}, we show the phase-folded ELODIE and TESS data after being subtracted from their corresponding linear trend and GP components. In Fig.~\ref{fig:corner_plot}, we show the corner plot of the main fitted parameters.

\subsection{Apsidal precession}

As shown in Sect.~\ref{sec:joint_fit}, the RVs of HD~118203 show a long-term linear trend that might be caused by an outer massive companion. This potential companion could perturb the orbit of HD~118203~b and cause apsidal precession. Determining whether the orbit of HD~118203~b precesses is of crucial importance to be able to interpret the possible links between the eccentric orbital motion of HD~118203~b and the photometric variations of its host star. We used the RVs to study whether there is any change in the argument of the periastron of the HD~118203~b orbit. To that end, we modelled the RVs following the procedure described in Sect. \ref{sec:joint_fit}, but parametrizing the argument of the periastron as $w_{0}$ + $dw \times (t-t_{0})$, being $w_{0}$ and $dw$ free parameters, $t$ the observing times, and $t_{0}$ the dataset mid-point. We obtain $w_{0}$ = 152.7 $\pm$ 3.1 deg and $dw$ = -0.01 $\pm$ 0.15 deg. Hence, being $dw$ consistent with zero, we conclude that the orbit of HD~118203~b does not significantly precesses throughout a 750-day time span.

\subsection{Stellar variability in TESS photometry}
\label{sec:var_tess}

The TESS photometry of HD 118203 shows a clear variability that can be seen in the four observed sectors (Fig.~\ref{fig:TESS_complete}). In order to analyse such photometric variations, we removed all the transit signatures of HD 118203 b. Given that the duration of the transits is $\simeq$5.7 hours, we masked all data points located three hours before and after each mid-transit time. 

\subsubsection{Search for periodic signals}
\label{sec:periodic_signals}

\begin{figure*}
    
    \centering
    \includegraphics[width=\textwidth]{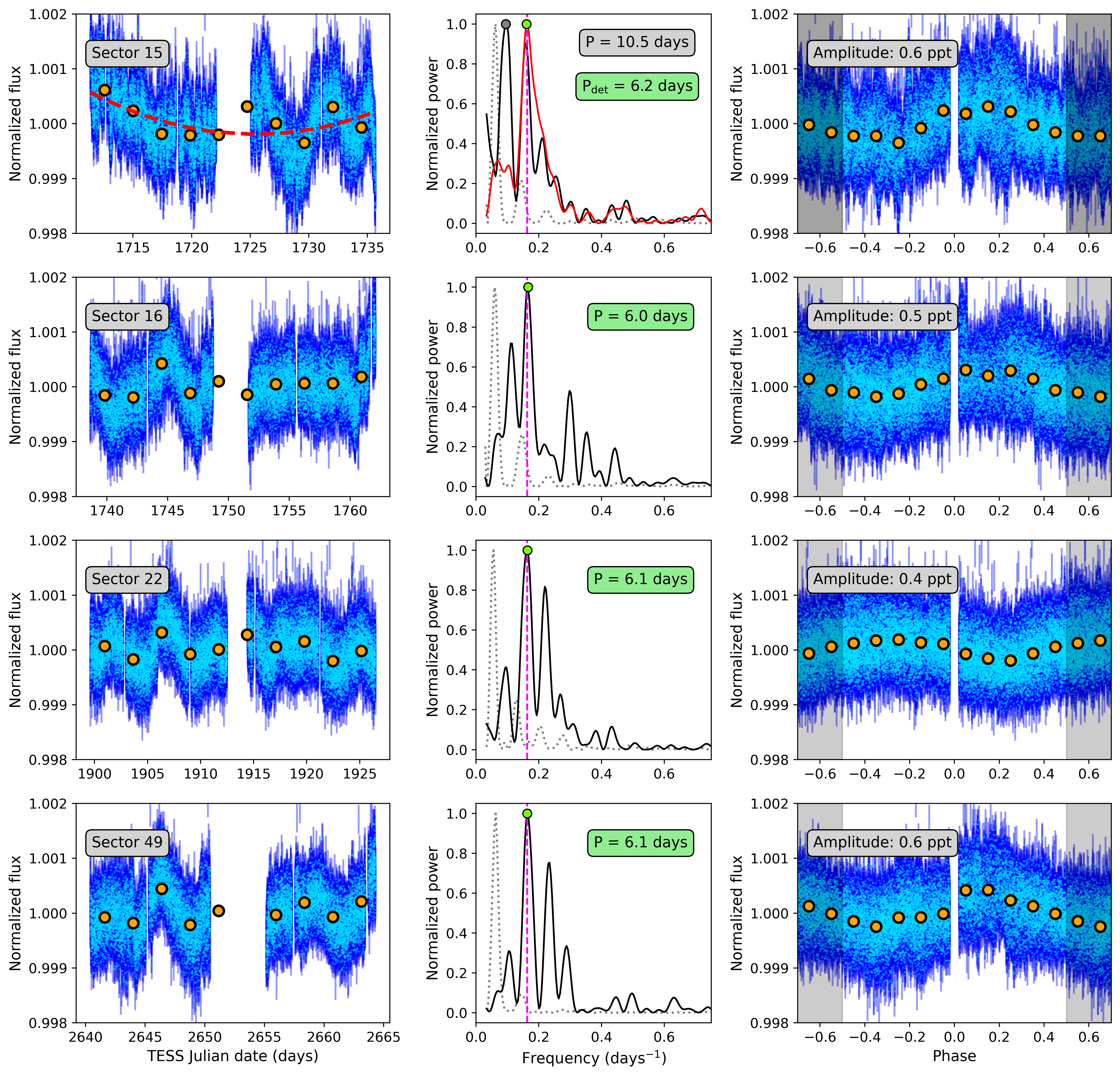}
    \caption{\texttt{GLS} periodograms of the TESS photometry time series and their window functions. \textit{Left panels:} TESS photometry time series of HD 118203. The red dashed line within the S15 panel indicates the degree 2 polynomial fit used to de-trend the data. The orange circles correspond to 1.8-day binned data. \textit{Centre panels:} \texttt{GLS} periodograms of the time series (solid black) and their window functions (dotted grey). In the S15 panel, the periodogram over the de-trended data is represented in red. The green circles and boxes indicate the maximum power frequencies and the vertical magenta dashed lines indicate the orbital period of HD 118203~b.  \textit{Right panels:} TESS photometry time series folded to the HD 118203~b orbital period as obtained from our global fit. The phase is referred to the planet's time of inferior conjunction.  The orange circles correspond to binned data of 10$\%$ the orbital phase.}
    \label{fig:tess_individual_periodograms}
\end{figure*}

We computed the generalized Lomb-Scargle periodogram \citep[\texttt{GLS};][]{2009A&A...496..577Z}\footnote{We used the \texttt{PyAstronomy} implementation \citep[][]{pya}, which is available at \url{https://github.com/sczesla/PyAstronomy}.} of the masked photometry in order to identify possible periodic sinusoidal-like signals. As a result, we obtain a maximum power period of 10.5 days in S15, 6.0 days in S16, 6.1 days in S22, and 6.1 days in S49, all of them with a False Alarm Probability (FAP)\footnote{All the FAPs in this work have been computed analytically following Eq. (24) of \citet{2009A&A...496..577Z}.} lower than 0.1$\%$. Thus, in three of the four observed sectors the TESS photometry shows a periodic sinusoidal-like signal with the same periodicity as the HD 118203~b orbit ($\simeq$6.1 days; see Table \ref{tab:bestfit}). Regarding S15, the first half of the light curve (just before the downlink gap) is dominated by a parabolic trend, but the second half shows a clear $\simeq$6-day periodicity, which stands out in the periodogram as the second highest peak. In order to know if this signal is significant, we detrended the S15 photometry for the parabolic trend, which is most likely the cause of the 10.5-day periodicity. To do so, we modelled the photometry with a second-order polynomial. The periodogram of the detrended photometry reveals a maximum power period of 6.2 days, also with FAP < 0.1$\%$. This result confirms that all the observed TESS sectors show a sinusoidal-like signal with a $\simeq$6.1-day periodicity, which coincides with the orbital period of the confirmed planet. We also ran the \texttt{GLS} over the complete TESS dataset and different combinations of sectors, and in all cases we detected the signal.

We studied whether the \texttt{GLS}-detected $\simeq$6.1-day signal describes a true periodic photometric variability, or if it could be an alias of the masked transits. To do so, we computed the \texttt{GLS} periodograms of the window functions of each sector. We find no peaks at the $\simeq$6.1-day periodicity detected within the original periodograms, and instead, we find maximum power periods of 16.1 days in S15, 15.8 days in S16, 18.0 days in S22, and 16.0 days in S49. Those periodicities correspond to 67$\%$ (i.e. two-thirds) of their corresponding sector lengths, which suggests that they could be related to the observing baselines. Alternatively, they could also be related to the beat period between the $\simeq$6.1-day transit gap separations and the length of each observing chunk before and after downlink. We filled the masked transit regions with mock data and checked whether the $\simeq$6.1-day periodicity and the 16-to-18 days periodicities of the window functions remain. To do so, we first performed a cubic spline interpolation over the TESS photometry filtered through a median filter with a 5-hour kernel size. Then, we filled the masked regions with 2-min cadence data generated according to the white noise properties of each sector; that is, we used Gaussian distributions with mean values centred on the interpolation models and standard deviations obtained from the flattened photometry. In Fig.~\ref{fig:mock_data_example}, we show an example of the generated mock photometry. The \texttt{GLS} periodograms of the filled datasets show maximum power periods of 6.1 days (S15), 6.0 days (S16), 6.1 days (S22), and 6.1 days (S49), all of them with FAP $<$ 0.1$\%$. Similarly, the \texttt{GLS} periodograms of the window functions of the filled datasets show the same maximum power periods as those of the window functions of the masked datasets (see Fig.~\ref{fig:wf_comparison}). This analysis confirms that the $\simeq$6.1-day signal found is not related to the data sampling (i.e. it is not an alias of the masked transits), and instead, it describes a true periodic photometric variability. In Fig.~\ref{fig:tess_individual_periodograms}, we show the sector-by-sector TESS photometry, the \texttt{GLS} periodograms of the original time series and their window functions, and the photometry folded in phase to the orbital period of HD 118203~b. 

\subsubsection{Persistence and evolution of the $\simeq$6.1-day signal}
\label{sec:persistence_evolution}

\begin{figure*}
    
    \centering
    \includegraphics[width = 0.47\textwidth]{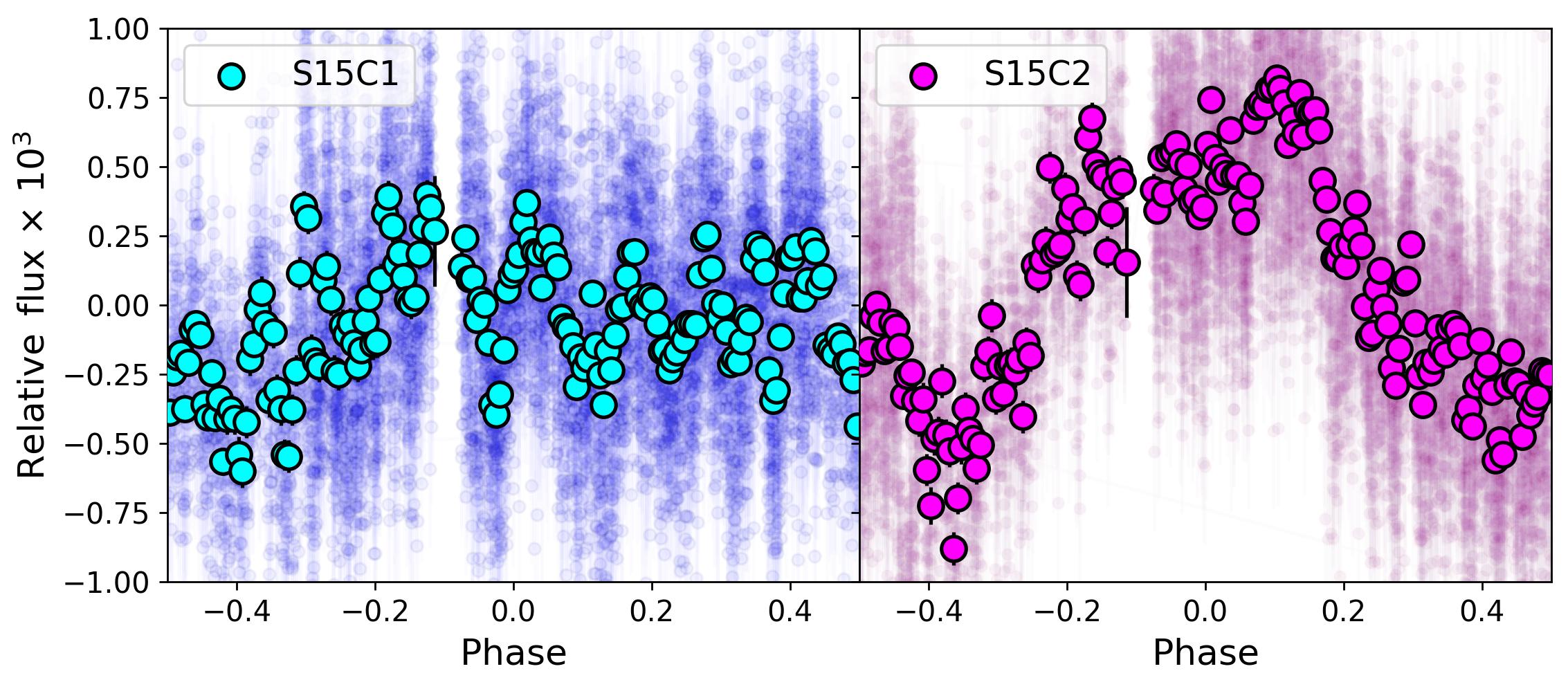}
    \includegraphics[width = 0.47\textwidth]{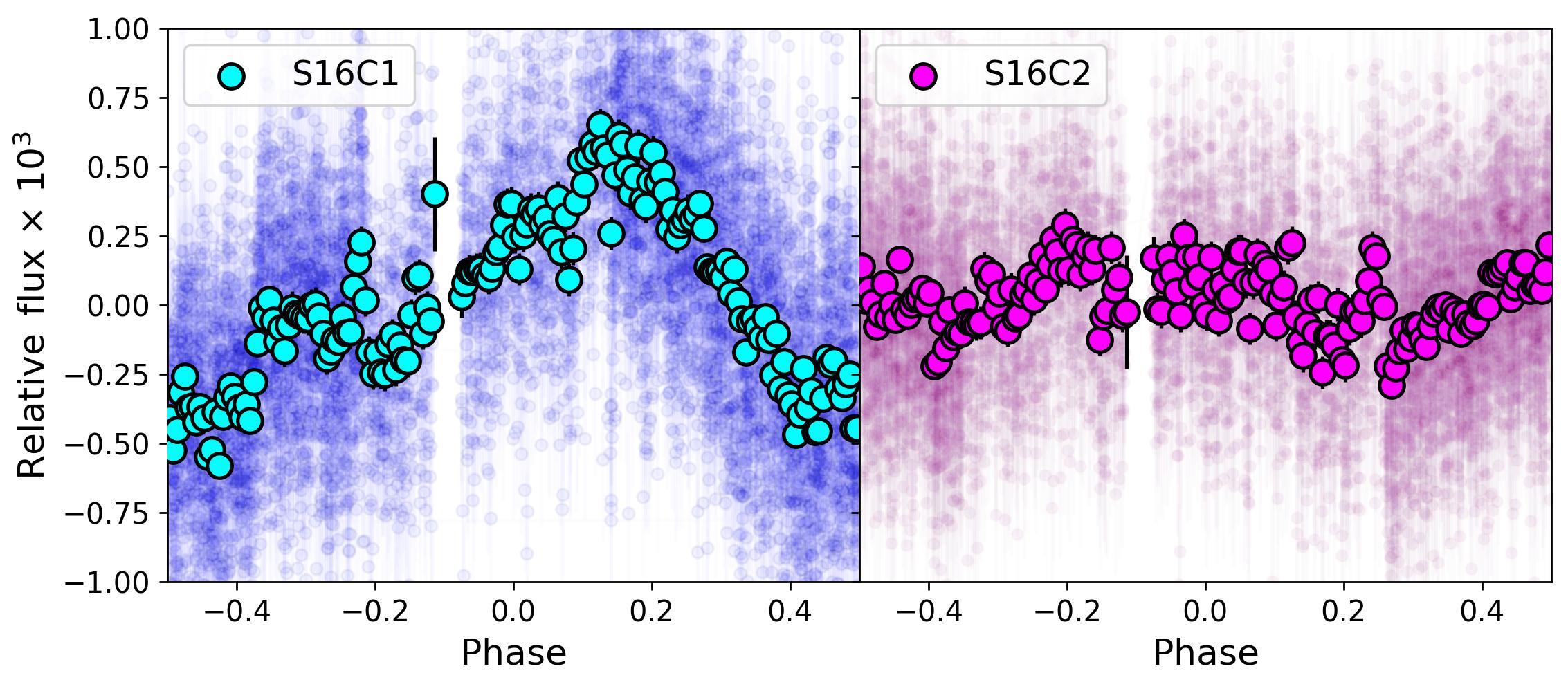}
    \includegraphics[width = 0.47\textwidth]{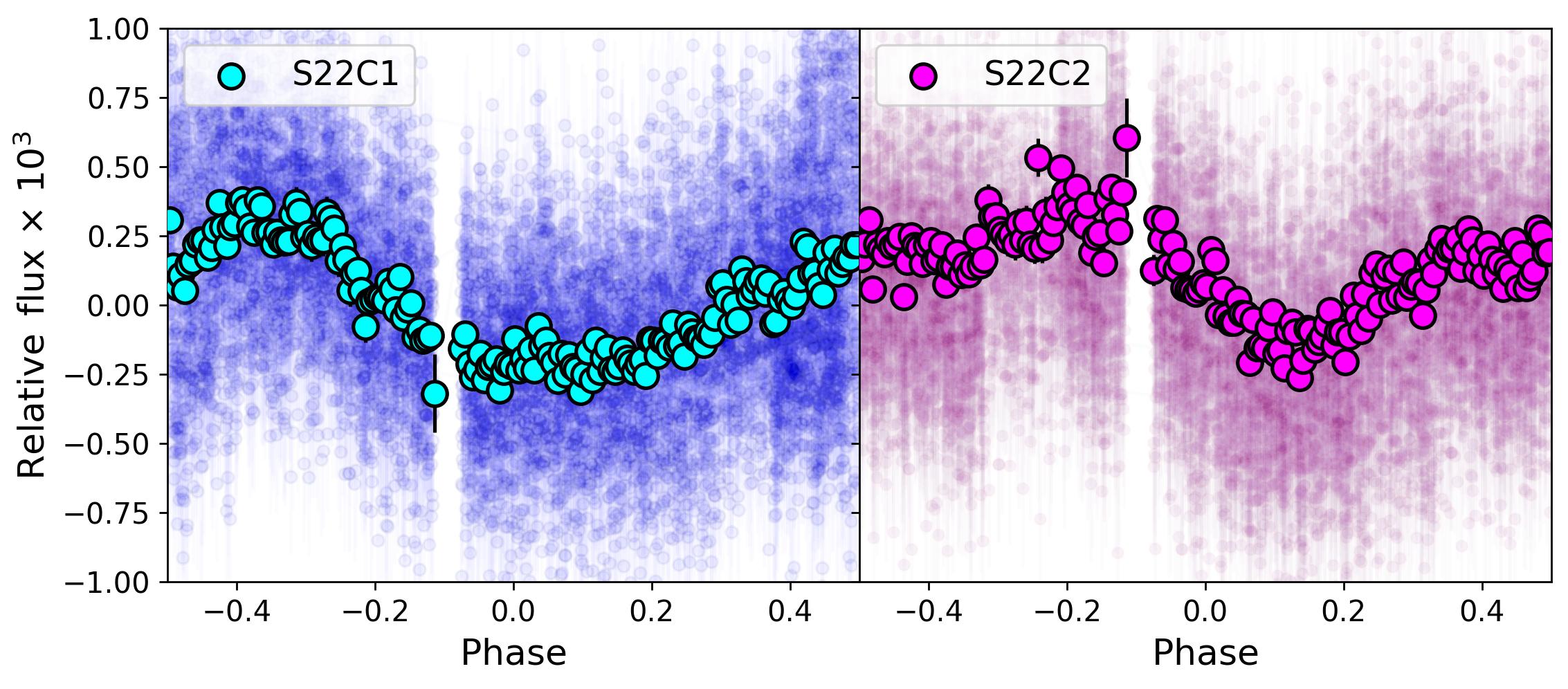}
    \includegraphics[width = 0.47\textwidth]{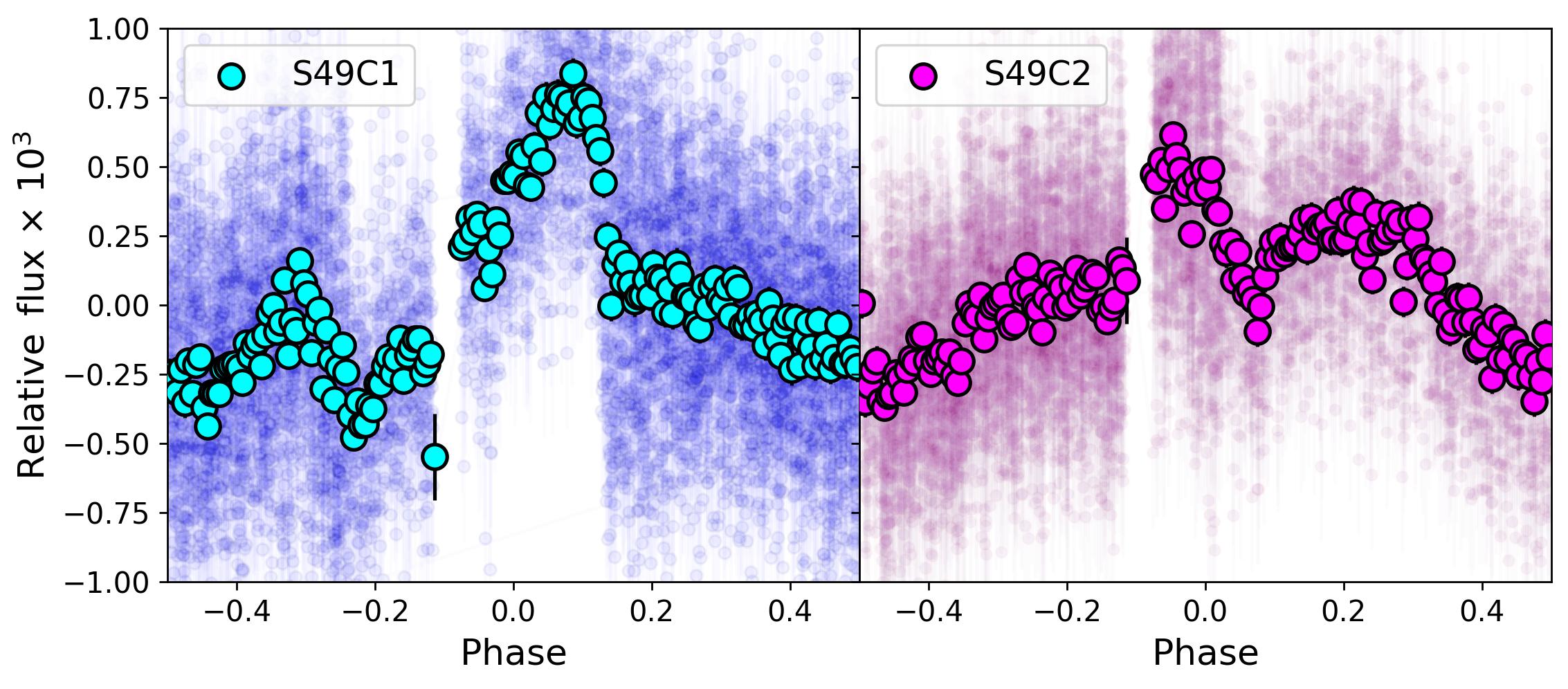}
    \caption{TESS PDCSAP split into eight chunks and folded in phase with the orbital period of HD~118203~b. Large circles correspond to 50-min binned data for better visualization of the photometric variations.}
    \label{fig:phase_with_planet}

\end{figure*}

We studied whether the $\simeq$6.1-day signal remains invariant or suffers changes from one orbit to another. To do so, we split the photometry of each sector into two chunks (i.e. before and after the downlink gap), which correspond to $\simeq$1.5 orbits. Hereinafter we refer to them as S$\mathcal{X}$C$\mathcal{Y}$, where $\mathcal{X}$ denotes the sector and $\mathcal{Y}$ denotes its corresponding chunk (i.e. 1 or 2). We repeated the \texttt{GLS} analysis over each chunk and detected the $\simeq$6.1-day signal in six of them. No significant signals were detected either in S15C1 or in S16C2. In Fig.~\ref{fig:phase_with_planet}, we plot each chunk folded in phase with the orbital period of HD 118203 b, in which we can appreciate how the $\simeq$6.1-day signal appears and disappears from one orbit to another in S15 and S16. 

Figure~\ref{fig:phase_with_planet} also reveals that the $\simeq$6.1-day signal experiences a strong and quick evolution. The magnitude of such evolution can be quantified in terms of the signal amplitude and shape. For example, the signal amplitude in S22C1 and S22C2 is decreased by $\simeq$50$\%$ with respect to that of S15C2 or S16C1. Also, the signal shape in S22 is inverted with respect to that of the other three sectors (i.e. in S15C2, S16C1, S49C1, and S49C2, the periodic transits of HD 118203 b occur before reaching the maximum flux emission, but in S22 the transits occur in the descendant region). Hence, in summary, we find that the $\simeq$6.1-day signal appears and disappears suddenly, and when it shows up it undergoes strong changes in both amplitude and shape. 

\subsubsection{Links with the orbital motion of HD 118203~b}
\label{sec:links}

\begin{table*}[]
\caption{Properties of the TESS photometric variations studied in Sect.~\ref{sec:links}.}

\renewcommand{\arraystretch}{1.2}
\setlength{\tabcolsep}{17.3pt}

\begin{tabular}{cccc|cc|cc}
\hline \hline
Sector (Chunk) & $f_{\rm inc}$ & $(f_{\rm inc})_{\rm vis}$ & $(f_{\rm inc})_{\rm hid}$ & $\phi_{\rm max}$ & $\phi_{\rm min}$ & R     & p-value \\ \hline
S15 (C2)       & 0.53      & 0.83              & 0.34              & +0.20        & -0.31        & -0.24 & 0.017   \\
S16 (C1)       & 0.56      & 0.88              & 0.38              & +0.25        & -0.36        & -0.18 & 0.070    \\
S22 (C1)       & 0.52      & 0.15              & 0.78              & -0.29        & +0.08        & -0.20 & 0.049    \\
S22 (C2)       & 0.52      & 0.46              & 0.55              & -0.08        & +0.20        & +0.33 & $7.3 \, \times \,10^{-4}$    \\
S49 (C1)       & 0.56      & 0.79              & 0.43              & +0.17        & -0.14        & +0.34 & $6.3 \, \times \,10^{-4}$     \\
S49 (C2)       & 0.54      & 0.70              & 0.41              & +0.30        & -0.36        & -0.13 & 0.18    \\ \hline
\end{tabular}
\tablefoot{$f_{\rm inc}$ is the orbital phase fraction in which the TESS photometric flux increases. $(f_{\rm inc})_{\rm vis}$ and $(f_{\rm inc})_{\rm hid}$ represent the same fraction but limited to the phases where a hypothetical co-rotating and small active region would be visible and hidden from Earth, respectively. $\phi_{\rm max}$ and $\phi_{\rm min}$ are the phase offsets between the subplanetary point and the maximum and minimum flux emission, respectively. R and p-value measure the correlations between the TESS flux derivative of HD~118203 and the orbital angular velocity of HD~118203~b.}
\label{tab:properties_variations}
\end{table*}

\begin{figure*}
     \centering

     \includegraphics[width = 5.62 cm]{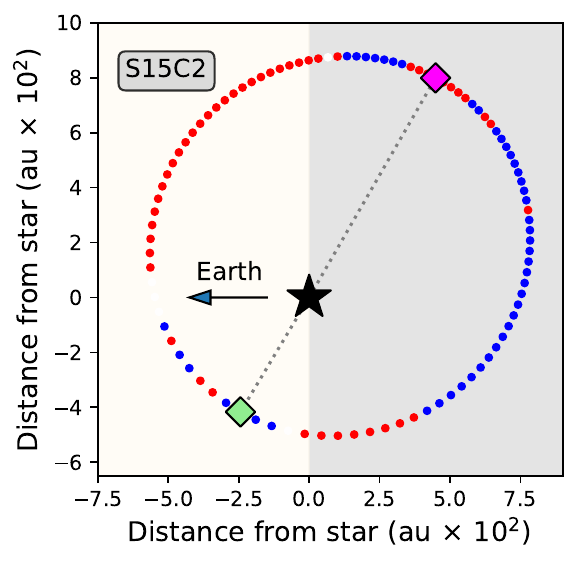}
     \includegraphics[width = 5.8 cm]{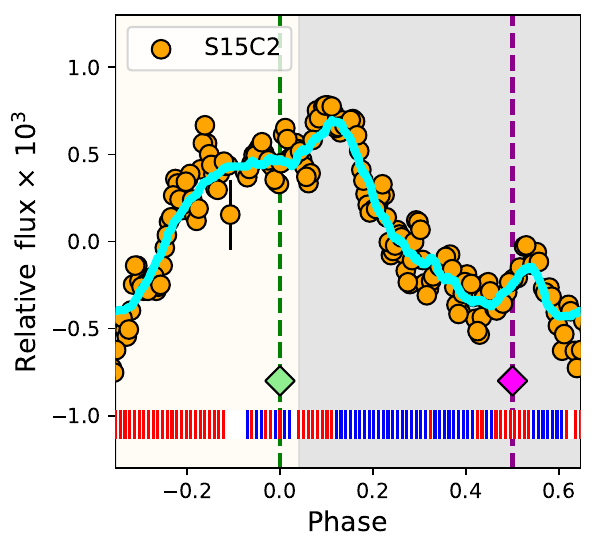}
     \includegraphics[width = 5.8 cm]{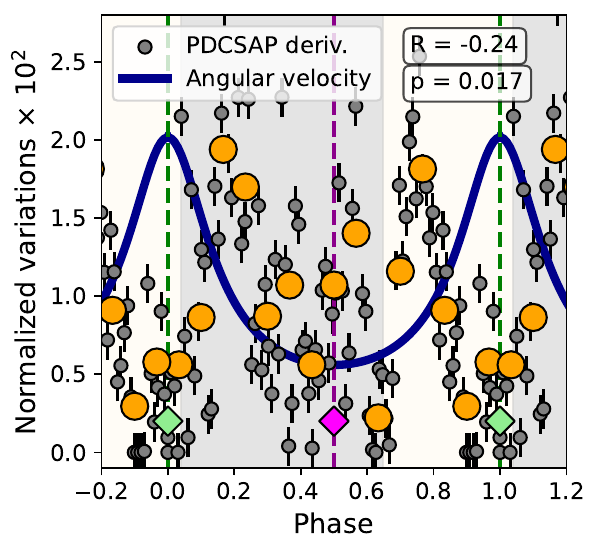}

     \includegraphics[width = 5.62 cm]{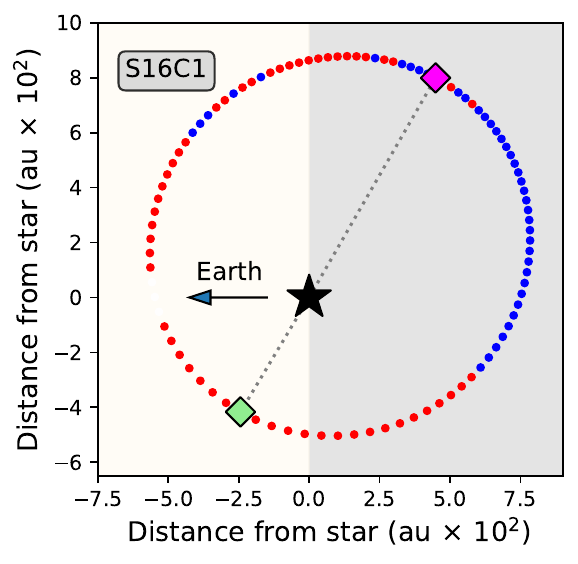}
     \includegraphics[width = 5.8 cm]{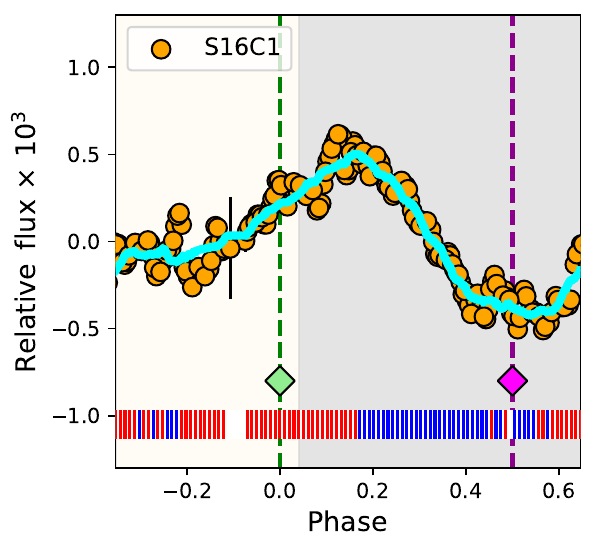}
     \includegraphics[width = 5.8 cm]{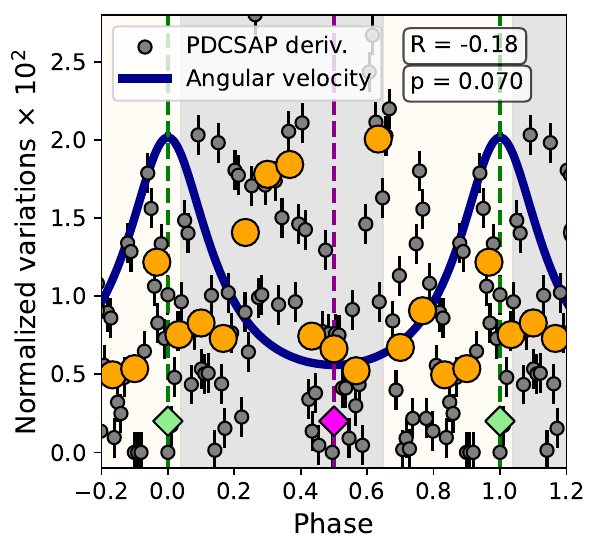}

     \includegraphics[width = 5.62 cm]{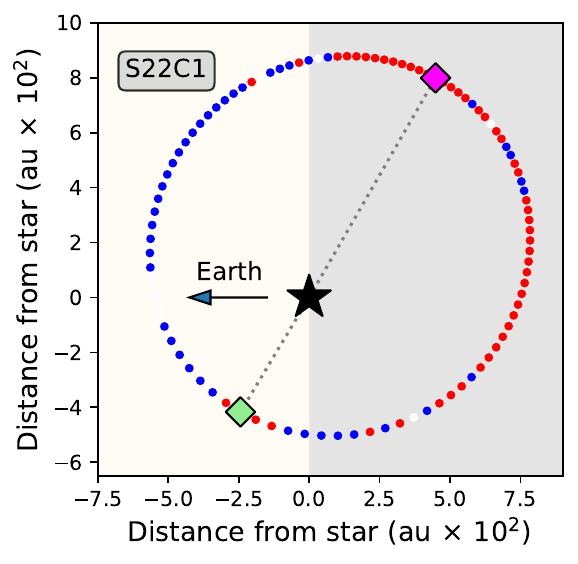}
     \includegraphics[width = 5.8 cm]{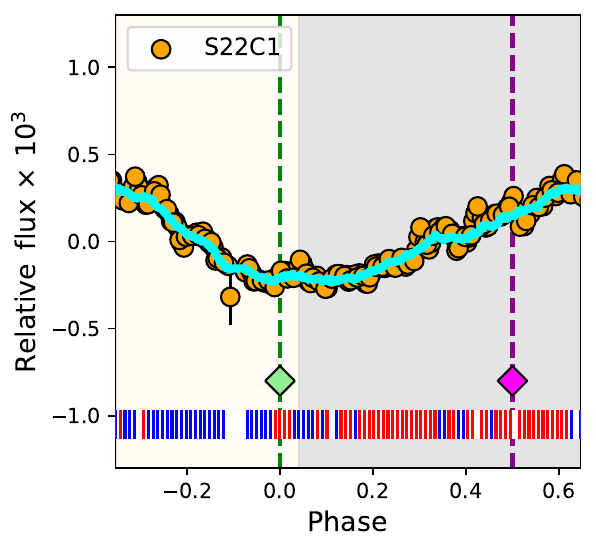}
     \includegraphics[width = 5.8 cm]{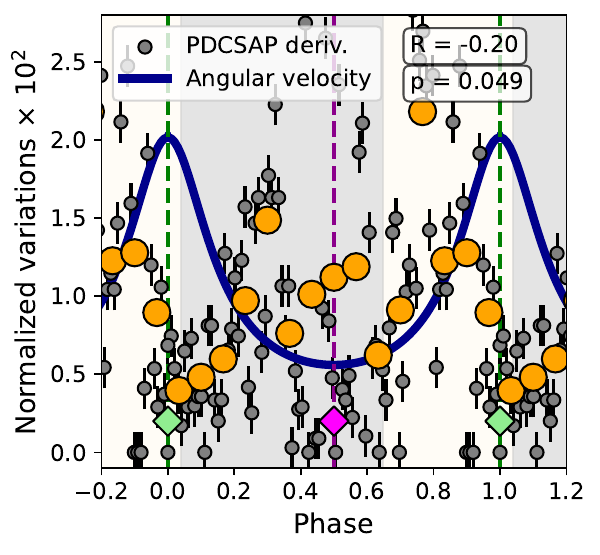}

    \caption{Links with the orbital motion of HD~118203~b. \textit{Left panels:} Orbital path of HD~118203~b, which follows the anticlockwise direction. The circles represent the location of HD~118203~b every $\simeq$1.5 hours and are coloured in red, blue, or white, depending on whether the photometric flux increases, decreases, or remains stable during that time lapse, respectively. \textit{Centre panels:} TESS PDCSAP photometry of HD~118203 folded in phase with the orbital period of HD~118203~b and binned with $\simeq$5$\%$ phase bins. The cyan line corresponds to the filtered photometry through a median filter with a kernel size of 701 cadences. The blue, red, and white vertical lines indicate whether the photometry increases, decreases, or remains stable in time lapses of $\simeq$1.5 hours. \textit{Right panels}: Comparison between the derivative of the phase-folded PDCSAP photometric variations of HD~118203 and the angular velocity of HD~118203~b. The orange circles correspond to binned data of 7$\%$ the orbital phase. \textit{In all panels:} The green and magenta squares represent the periapsis and apoapsis of the orbit, respectively, and the white and grey backgrounds represent the orbital regions in which a hypothetical stellar co-rotating active region would be visible and not visible from Earth, respectively.}

    \label{fig:main_plot}
        
\end{figure*}

\begin{figure*}

     \includegraphics[width = 5.62 cm]{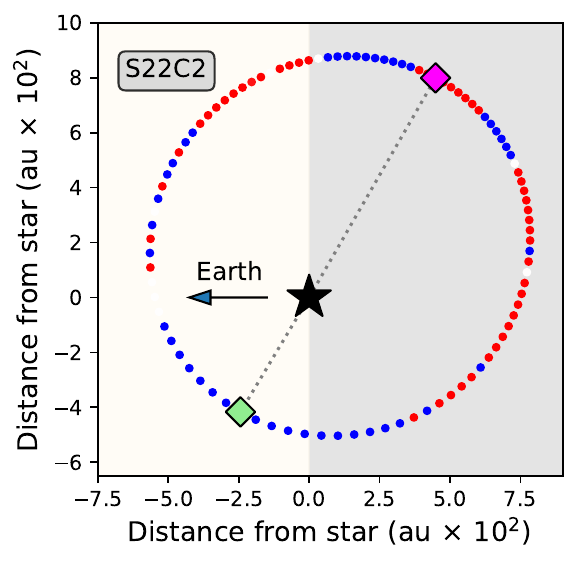}
     \includegraphics[width = 5.8 cm]{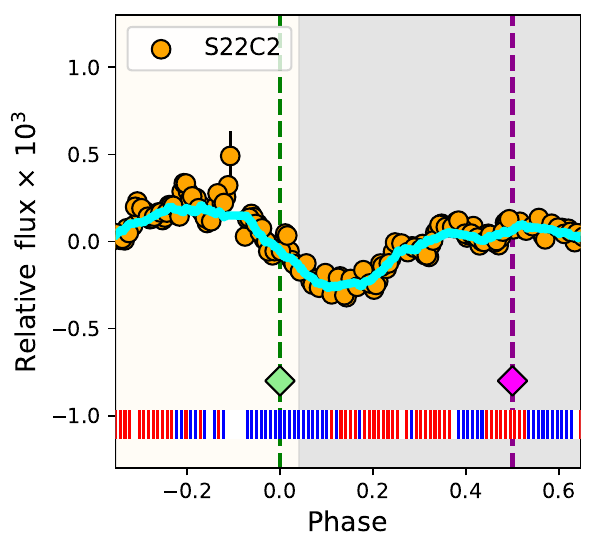}
     \includegraphics[width = 5.8 cm]{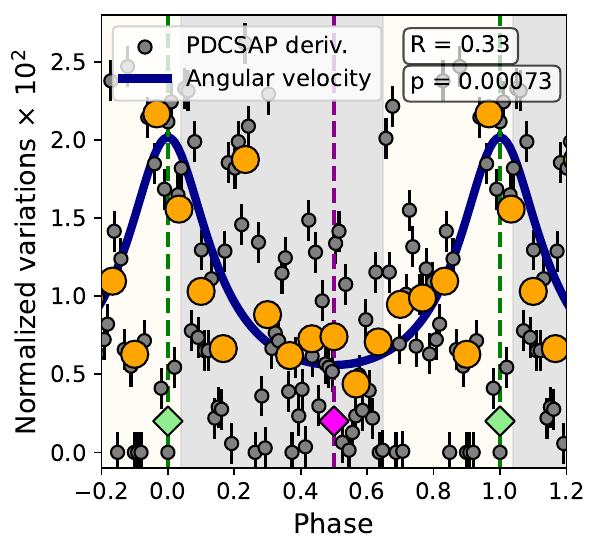}

     \includegraphics[width = 5.62 cm]{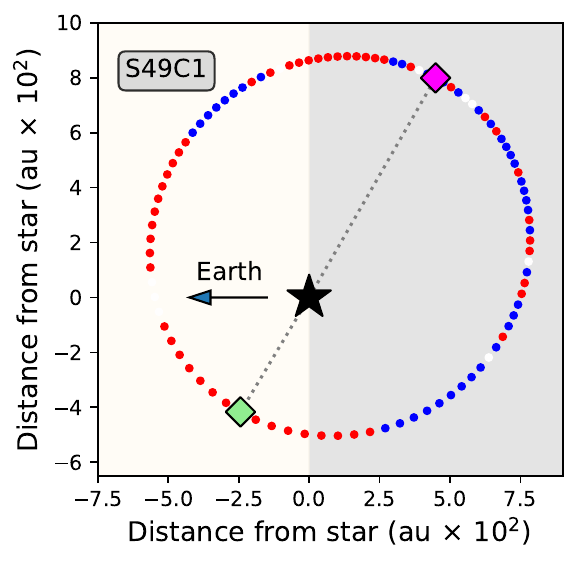}
     \includegraphics[width = 5.8 cm]{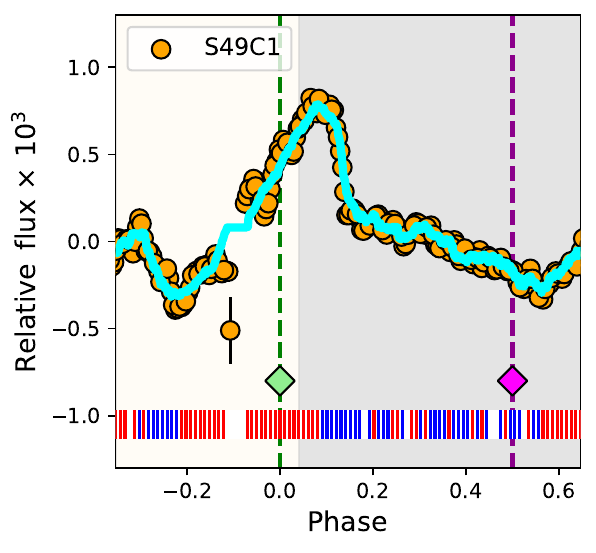}
     \includegraphics[width = 5.8 cm]{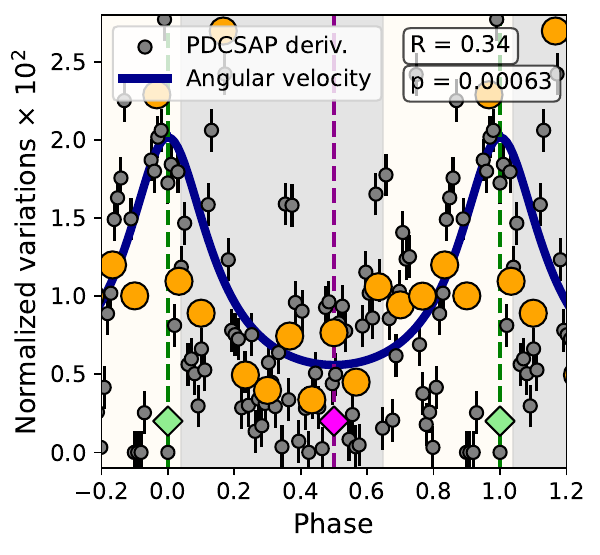}

     \includegraphics[width = 5.62 cm]{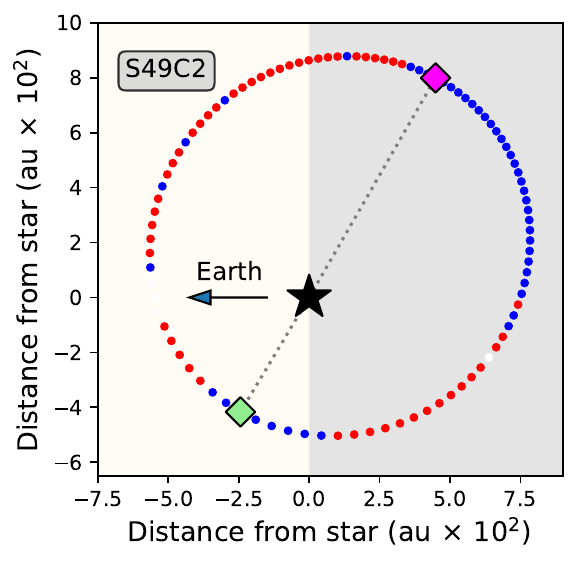}
     \includegraphics[width = 5.8 cm]{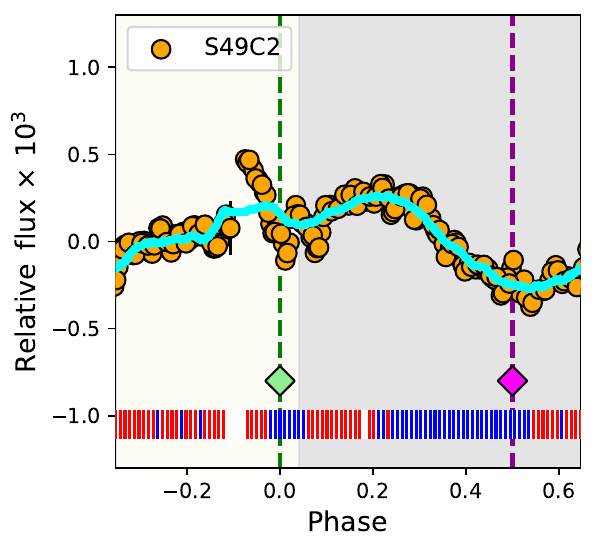}
     \includegraphics[width = 5.8 cm]{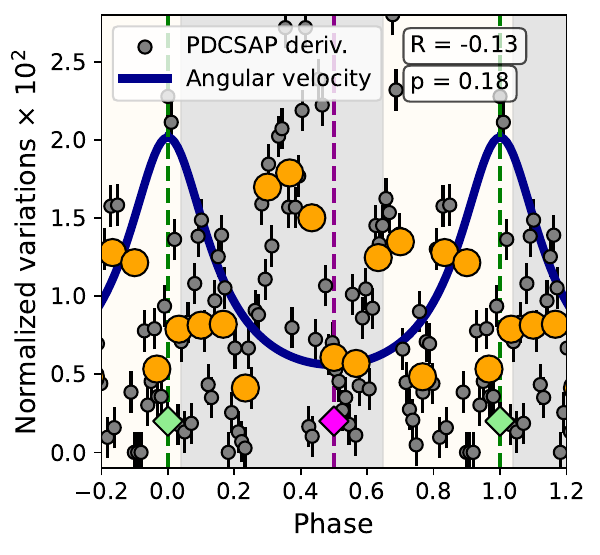}

     \caption{Continuation of Fig~\ref{fig:main_plot}.}

     \label{fig:main_plot_2}

\end{figure*}

We studied whether the $\simeq$6.1-day photometric signal is linked to the orbital motion of HD~118203~b. That is, we compared the observed photometric variations with the location and angular velocity of HD~118203~b throughout its eccentric orbit. 

We divided the complete phase-folded PDCSAP photometry into 100 bins of $\simeq$1.5 hours long, for which we estimated the flux derivative. In order to mitigate the short-term variations induced by the photometric scatter, we previously filtered the phase-folded photometry through a median filter with a kernel size of 701 cadences (i.e. $\simeq$10$\%$ of the orbital phase) and then computed the flux differences of the filtered photometry. In Table~\ref{tab:properties_variations}, we include the orbital phase fraction in which the photometric flux increases ($f_{\rm inc}$)\footnote{We note that in the transit regions, there are three cadences in which we do not have information on the flux derivative. In those cases, we considered that the flux increases or decreases when there are at least four cadences before and after with the same derivative sign; that is, in S16C1, S22C1, S49C1, and S49C2.}. We find that those fractions roughly correspond to half an orbit. In Table~\ref{tab:properties_variations} we also include the orbital fraction of increasing photometry limited to the phases where a hypothetical co-rotating and small active region would be visible $(f_{\rm inc})_{\rm vis}$ and hidden $(f_{\rm inc})_{\rm hid}$ from Earth. In this case, we find a clear imbalance between the increasing and decreasing regions. In S15C2, S16C1, S49C1, and S49C2, the flux keeps increasing throughout 70-90$\%$ of the time that the hypothetical co-rotating active region would be visible from Earth. On the contrary, in S22C1 the flux decreases during 85$\%$ of the visible phase, and in S22C2 the increasing and decreasing regions last similar.  We also computed the phase offset between the subplanetary point (i.e. time of mid-transit) and the maximum ($\phi_{\rm max}$) and minimum ($\phi_{\rm min}$) flux emission based on the fitted orbital parameters and the filtered photometry. The results are included in Table~\ref{tab:properties_variations}. We find that no flux extreme coincides with the subplanetary point, and instead, there are offsets between -0.35 and 0.25 orbital phases. In order to visualize the aforementioned results, in the left panels of Figs.~\ref{fig:main_plot} and \ref{fig:main_plot_2} we represent the orbital motion of HD 118203~b based on the fitted parameters in Sect.~\ref{sec:joint_fit}, and in the centre panels of the same figures we show the phase-folded binned and filtered photometry. 

Based on Kepler's second law, we computed the angular velocity of HD 118203~b as:

\begin{equation}
    w(r) = \frac{d\theta}{dt} = \frac{\pi}{P_{\rm orb} r^{2}} \left( d_{a} + d_{p} \right) \sqrt{d_{a} d_{p}}
\end{equation}

where $r$ is the distance between the star and the planet, and $d_{p}$ and $d_{a}$ are the distances between the star and the periapsis and apoapsis, respectively. We studied whether this quantity is correlated with the absolute photometric derivative. A certain degree of correlation would be expected in a scenario in which the active regions are moving over the stellar surface synchronized with the eccentric orbital motion of HD 118203 b. We note that there is a geometric factor that also affects the flux derivative. However, we cannot take it into account since it depends on the unknown size, location, evolution, and amount of active regions generating the variations. Hence, we warn the reader to just take this analysis as a first approach to quantify a possible link between the eccentric orbital motion of this planet and the photometric variations generated by hypothetical active regions co-rotating with it. In order to make the comparison, we previously normalized each quantity through the total accumulated variations, which we estimated by using the trapezoidal rule. We quantified the existence of possible correlations by means of the Pearson product-moment correlation coefficient and its associated p-value \citep{doi:10.1080/00031305.1988.10475524}, which we include in Table \ref{tab:properties_variations}. We find a low degree of correlation (R < 0.3) for S15C2, S16C1, S22C1, and S49C2, and we find a moderate degree of correlation (0.30 < R < 0.50) for S22C2 and S49C1, with R coefficients of 0.33 and 0.34, and p-values of $7.3\,\times\,10^{-4}$ and $6.3\,\times\,10^{-4}$, respectively. In the right panels of Figs.~\ref{fig:main_plot}~and~\ref{fig:main_plot_2}, we show the absolute photometric derivatives and the angular velocity of HD~118203~b resolved in phase, where we can visualize the moderate correlation found in S22C2 and S49C1 as well as the lack of correlation in the remaining sectors.

\subsection{Stellar variability in ELODIE spectroscopic data}
\label{sec:variability_ELODIE}

\begin{figure*}
    \centering
    \includegraphics[width=\textwidth]{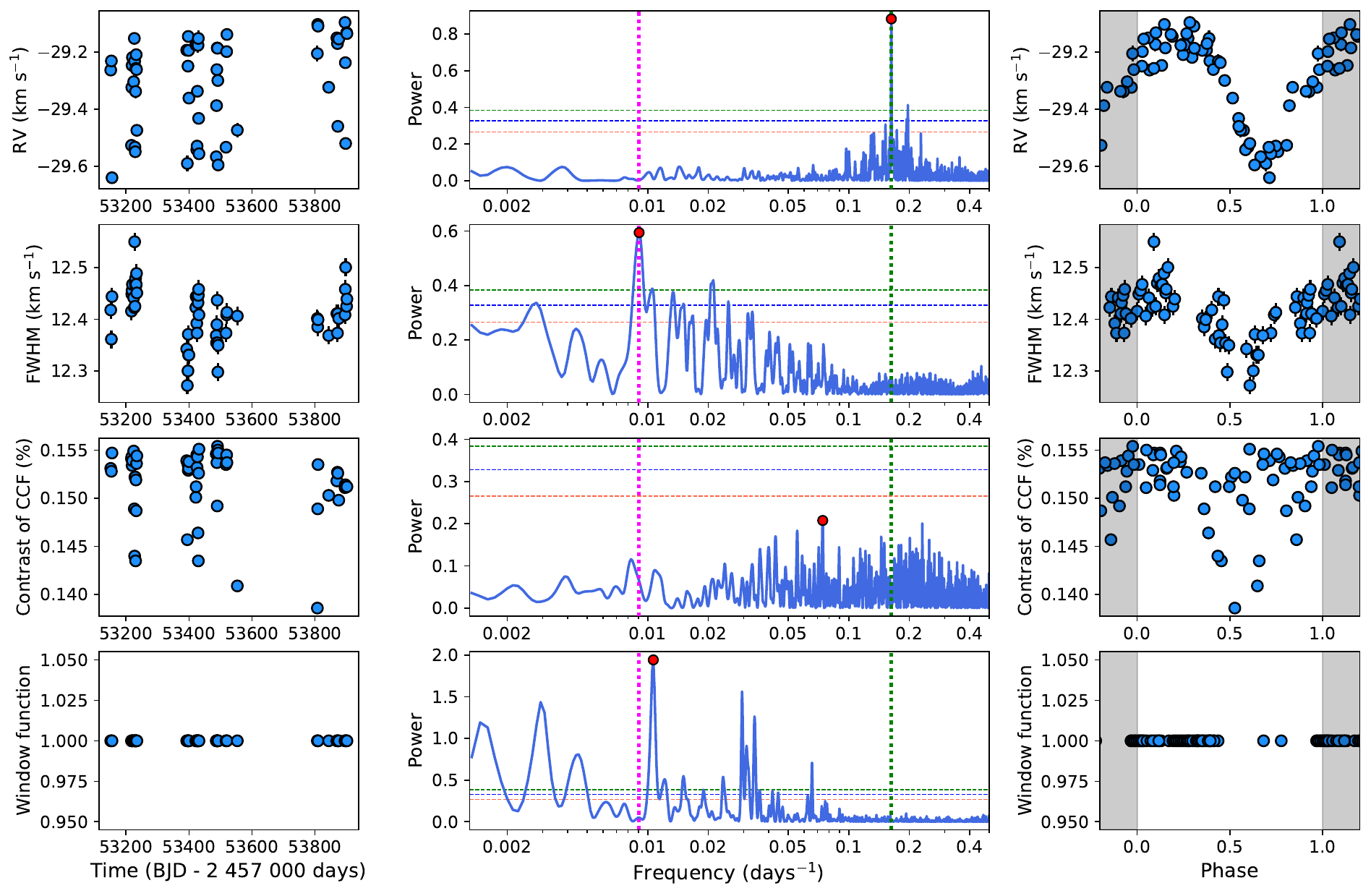}
    \caption{\texttt{GLS} periodograms of the ELODIE time series and window function. \textit{Left panels}: Time series of the ELODIE RVs, activity indicators described in Sect.~\ref{sec:obs_elodie}, and window function of the observations. \textit{Centre panels}: \texttt{GLS} periodogram of the time series and the window function. The red circle indicates the maximum power frequencies. The green dotted vertical lines highlight the $\simeq$6.1-day orbital period of HD~118203~b. The magenta dotted vertical lines indicate the $\simeq$110-day periodicity found in the ELODIE FWHMs. The horizontal dotted lines correspond to the 10 (orange), 1 (blue), and 0.1$\%$ (green) FAP levels. \textit{Right panels}: ELODIE time series and window function folded to the maximum power periods.}
    \label{fig:gls_to_ELODIE}
\end{figure*}

We computed the \texttt{GLS} periodogram of the ELODIE RVs and activity indicators described in Sect.~\ref{sec:obs_elodie}. The RVs periodogram shows a maximum power period of 6.1 days with a FAP of $10^{-20}\%$ that corresponds to the orbital period of HD~118203~b. The FWHMs periodogram shows a maximum power period of 110.3 days with a FAP of $10^{-6}\%$ that suggests the existence of a periodic activity signal. This signal has no counterpart either in the RVs or in the contrasts of the CCFs, the latter being devoid of any signal with FAP $<$ 10$\%$. We also computed the \texttt{GLS} periodogram
of the window function of the observations in order to determine whether the detected peaks could be related to the sampling of the data. We find a maximum power period of 93.8
days, which is relatively close to the periodicity detected in
the FWHMs. In Fig.~\ref{fig:gls_to_ELODIE}, we show the computed periodograms together with the ELODIE time series folded in phase to their corresponding maximum power periods.

\subsection{Stellar variability in ASAS-SN photometry}
\label{sec:variability_asassn}

\begin{figure*}
    \centering
    \includegraphics[width=\textwidth]{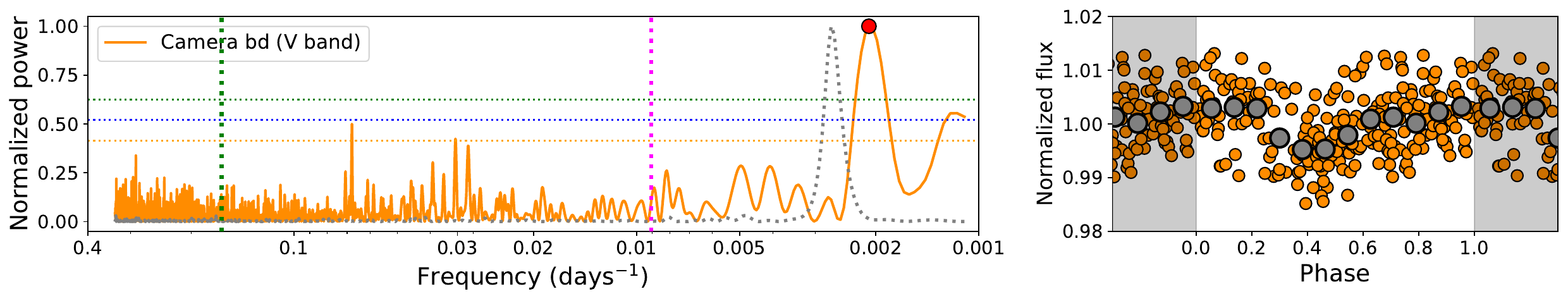}
    \caption{\texttt{GLS} periodogram of the ASAS-SN photometry acquired by the camera bd (solid orange) and its window function (dotted grey). The red circle indicates the maximum power frequency. The green dotted vertical line highlights the $\simeq$6.1-day orbital period of HD~118203~b, and the magenta dotted vertical line indicates the $\simeq$110-day periodicity found in the ELODIE FWHMs. The horizontal dotted lines correspond to the 10 (orange), 1 (blue), and 0.1$\%$ (green) FAP levels. The right panel illustrates the phase-folded photometry to the maximum power period. The grey data points correspond to a 35-day binning.}
    \label{fig:asas-sn}
\end{figure*}

We computed the \texttt{GLS} periodograms of the ASAS-SN photometry acquired by the four cameras. The periodograms of three of them show no significant periodic signals; that is, there are no peaks with FAP $<$ 10$\%$. The periodogram of the camera \textit{bd}, however, shows a peak at 477 days with a FAP of $\simeq$ $10^{-5}$$\%$, but no peak appears at the 110.3-day signal found in the ELODIE FWHMs.  We also computed the \texttt{GLS} periodogram of the window function. It shows a maximum power period of 366 days, which is possibly related to a yearly alias due to the seasonal visibility of the target. In Fig.~\ref{fig:asas-sn}, we show the computed periodograms together with the ASAS-SN photometry folded in phase to its maximum power 477-day period.

\section{Discussion}
\label{sec:discussion}
\subsection{Origin of the $\simeq$6.1-day TESS photometric signal}

Close-in giant planets can modify the activity levels of their host stars through MSPIs \citep[][]{2000ApJ...533L.151C,2003ApJ...597.1092S,2005ApJ...622.1075S,2008ApJ...676..628S,2008A&A...482..691W,2009EM&P..105..373P,2011ApJ...741L..18P,2015ApJ...811L...2M,2019NatAs...3.1128C}. The commonly considered criterion to confirm those interactions consists of detecting a periodic activity signal that matches the orbital period of the hosted planet, as well as another periodic signal induced by the rotation of the star. In this work, we have found evidence of periodic stellar variability in the TESS photometry of HD 118203 that matches the orbital period of its hosted giant planet HD 118203~b (i.e. $\simeq$6.1 days; Sect.~\ref{sec:var_tess}). We searched for possible rotation-induced activity in the ELODIE activity indicators (Sect.~\ref{sec:variability_ELODIE}) and ASAS-SN photometry (Sect.~\ref{sec:variability_asassn}) and found significant signals at $\simeq$110 and $\simeq$477 days, respectively. However, given the disparity of results in both datasets, the non-repeatability of the signals found, and the relative closeness of the ELODIE periodicity to the highest peak of the window function periodogram, we consider that we do not have enough evidence to confidently confirm the rotation period of HD~118203. 

With a mass of $\simeq$1.3 $\rm M_{\odot}$, HD 118203 must have been a F-type star during its main sequence (MS) phase. Such a mass coincides with the Kraft break \citep{1967ApJ...150..551K}. Stars with lower masses are known to have large convective envelopes that provoke angular momentum loss due to magnetized stellar winds. In contrast, stars with higher masses are known to have very thin convective envelopes, so they can keep rotating quickly during the MS phase. As stars leave the MS, their cores contract and their envelopes expand, resulting in an increase in their moments of inertia that leads to a decrease in the surface rotation rate as the star evolves across the subgiant branch. Besides, stars with masses above the Kraft break develop convective envelopes so that wind-driven angular momentum loss affects both MS rotational regimes in evolved stages \citep[e.g.][]{2013ApJ...776...67V}. This has been empirically shown by \citet{2012A&A...548L...1D}, who found that subgiants with masses both above and below the Kraft break show rotation periods of several tens of days. Therefore, whether HD 118203 underwent significant magnetic braking during its MS phase, or remained rotating rapidly because of a thin convective envelope, both models and theory indicate that its rotation period would have been increased during its subgiant phase.

Another factor that might also affect the rotation velocity of this star is the presence of a close-in massive giant planet. However, the energy dissipation due to tidal forcing is much lower in stars than in their hosted planets \citep[e.g.][]{Fabrycky_Tremaine_2007,Correia_etal_2011,Beauge_Nesvorny_2012}. This has been observationally proved for G-type stars with tightly orbiting giant planets, which show similar rotation periods to those of G-type stars without close-in companions \citep{2013PASP..125..989A}. F-type stars, however, are less studied in this regard, but we know they should be more prone to reach a synchronization state since radiative envelopes dissipate more energy than convective envelopes \citep[e.g.][]{2011A&A...529A..50L}, and, as discussed above, they are much less slowed down by magnetized winds. Interestingly, to date, the only star thought to be tidally locked to the orbit of its hosted hot Jupiter is an MS F-type star, $\tau$ Boo \citep{1997ApJ...474L.115B,2000ApJ...529L..41H}. Therefore, the natural question that arises here is whether HD~118203 may be a new case of a tidally locked star to its hosted hot Jupiter. If so, the tidal locking configuration must have been strong enough to avoid the influence of the braking forces experienced by the star when ascending the subgiant branch. The tidal theory shows that the equilibrium state of planetary systems with gas giants in eccentric orbits is not a synchronous state, but rather a pseudo-synchronous state that depends on the orbital eccentricity \citep{1981A&A....99..126H,Correia_etal_2011}. According to Eq. (42) in \citet{1981A&A....99..126H}, the pseudo-synchronization in the HD 118203 planetary system corresponds to a rotation period of 3.73 days. Such pseudo-synchronized rotation is expected to have occurred quickly for the hosted giant planet HD 118203~b (i.e. a few million years), but much more slowly for the host star HD 118203 (in the case that this rotational state was actually reached). Therefore, being HD 118203 a subgiant star, either coming from an MS star with a thick or narrow convective envelope, we expect that it underwent the well-studied rotation period increase when ascending the subgiant branch. In a hypothetical and unusual case in which the rotation of HD 118203 could have been dominated by the tidal forces exerted by its hot Jupiter during its MS phase, the star would be pseudo-synchronized at 3.73 days, being a hypothetical 6.1-day rotation a non-equilibrium state for this system. Hence, although we do not have observational evidence of the rotation period of HD 118203, we argue that it being 6.1 days is very unlikely, which makes MSPIs the most likely scenario to explain the observed TESS activity signal. This argument agrees with literature estimates of the spectroscopic projected rotational velocity in case the stellar rotation axis is orthogonal to the line of sight. That is, \citet{2006A&A...446..717D} and \citet{2017AJ....153...21L} computed values of 5 and 7 km$\rm \,s^{-1}$, respectively, while a projected velocity of 17 km$\rm \,s^{-1}$ would be required for the rotation of the star to be the 6.1-day signal found. In the case of a pseudo-synchronized 3.73-day rotation, a 27 km$\rm \,s^{-1}$ projected rotational velocity would be required. In the following, we discuss how the TESS signal properties unveiled in Sects. \ref{sec:persistence_evolution} and \ref{sec:links} match the MSPIs and rotation scenarios. 

\subsection{TESS signal properties}

In Sect.~\ref{sec:persistence_evolution}, we found that the $\simeq$6.1-day photometric signal is not a persistent phenomenon (see Fig.~\ref{fig:phase_with_planet}). Interestingly, this is an expected behaviour for MSPIs. \citet{2008ApJ...676..628S} found evidence of synchronous stellar activity in HD 179949 and $\upsilon$ And during 75$\%$ of the monitored time, while in the remaining 25$\%$ only the rotation of the star showed up. This on/off nature of MSPIs had been previously predicted by the models of \citet{2007astro.ph..2530C}. The authors found that the complex nature of the multipole fields may cause MSPI-induced activity not to repeat exactly from one orbit to another, and sometimes it can even disappear completely. We note, however, that while being expected for MSPIs, this behaviour does not completely rule out the rotation of the star as the source of the variability. Stellar rotation is a periodic phenomenon, but the rotation-induced activity signals are quasiperiodic since the active regions move on the stellar surface and appear and disappear throughout the magnetic cycle timescale. The typical stellar magnetic cycles last several years \citep[e.g.][]{2016A&A...595A..12S}, but some stars have also been found to have a quick spot evolution, showing variations throughout time scales of a few days \citep[e.g.][]{2019ApJ...871..187N}. Therefore, we consider that the strong changes in amplitude and shape experienced by the $\simeq$6.1-day TESS signal are most likely explained by MSPIs, although we cannot completely discard the possibility that they could be caused by the rotation of the star with quick-evolving spots.

In Sect.~\ref{sec:links}, we studied the possible existence of links between the $\simeq$6.1-day photometric signal and the orbital motion of HD~118203~b. We found that the flux variations span the complete orbit, which suggests that the total active area generating those variations occupies a considerable amount of the stellar surface. Hence, the observed photometric properties do not allow us to identify a link with the orbital location of the planet under the assumption of a small co-rotating spot on the stellar surface. However, we also found that in five of the six analysed chunks the flux mostly increases (S15C2, S16C1, S49C1, and S49C2) or decreases (S22C1) when such a hypothetical big active area would be visible from Earth. Hence, if we consider an MSPI scenario, the photometry of S15C2, S16C1, S49C1, and S49C2 would be best explained by the presence of bright and extensive spots co-rotating with HD~118203~b, while in S22 such co-rotating spots would be dark. This possible alternation caused by MSPIs has been previously noticed by \citet{2008A&A...482..691W}. Previous works reporting MSPIs also found that the stellar activity extremes tend to be shifted with respect to the subplanetary point, which is interpreted as the co-rotating active regions advancing or delaying with respect to the planet location \citep[e.g.][]{2008ApJ...676..628S}. Depending on the system, those offsets can vary significantly  (e.g. \citealp{2019NatAs...3.1128C} found a 0.46 phase offset for $\upsilon$ And, a 0.92 phase offset for HD 189733, and offsets in between for other two systems). Interestingly, we have found similar offsets for HD~118203.

In Sect.~\ref{sec:links}, we also studied whether the eccentric orbital motion of HD~118203~b has an imprint in the observed TESS variability. To date, only two eccentric systems had been reported to possibly have MSPIs, Kepler-432 \citep{2015ApJ...803...49Q,2015A&A...573L...5C} and  HD 17156 \citep{2015ApJ...811L...2M}. However, the periodic variability of Kepler-432 was found to be invariant during the whole photometric observations so that both MSPIs and tidal interactions could explain them \citep{2015ApJ...803...49Q}. Regarding HD~17156, the occasional nature of the activity enhancement makes it also compatible with accretion onto the star of material tidally stripped from the planet \citep{2015ApJ...811L...2M}. Therefore, HD~118203~b represents the best evidence of MSPIs in eccentric planetary systems, which makes it an ideal target to probe a possible eccentricity imprint in the observed photometric variations. We searched for a possible correlation between the stellar flux derivative ($\frac{df}{dt}$) and the planetary angular velocity ($\frac{d\theta}{d t}$), given that the projected area of active regions connected to, and co-moving with the planet, would change according to the eccentric orbital motion of the planet. Unfortunately, we could not confirm the existence of clear links between the photometric variations of HD~118203 and the eccentric orbital motion of HD~118203~b. 

\subsection{Orbital eccentricity as a possible booster of MSPIs}

As discussed before, given the closeness of this gas giant to its host star, and the significant orbital eccentricity, the rotation period of HD~118203~b is expected to be pseudo-synchronized at $\simeq$3.73 days. This prediction indicates that the eccentric orbit of this hot Jupiter could be responsible for a higher planetary magnetic moment than for a similar circular system. In particular, HD~118203~b  would have a $\simeq 40\%$ larger magnetic moment than if it was in a circular synchronized orbit. \citet{2005ApJ...622.1075S,2008ApJ...676..628S} found that circular planetary systems with hot Jupiters with $M_{\rm p}$sin(i) / $P_{\rm rot}$ $>$ 0.4 $M_{J}$ $\rm day^{-1}$ tend to undergo MSPIs, while below that value only 20 $\%$ of their studied systems showed MSPIs. For HD~118203~b, we estimate a $M_{\rm p}$ / $P_{\rm rot}$ of 0.61 $M_{J}$ $\rm day^{-1}$, while in a hypothetical circular orbit, the ratio would be 0.37 $M_{J}$ $\rm day^{-1}$. Hence, the estimated ratio for the eccentric case is well above the 0.4 $M_{J}$ $\rm day^{-1}$ value where MSPIs are expected to be found, while in a hypothetical circular orbit, the estimated ratio corresponds to a region of the parameter space where MSPIs are not expected to be always found. This suggests that the unusually high eccentricity of HD~118203~b could be critical for the generation of the potential MSPIs found.

\section{Summary and conclusions}
\label{sec:conclusions}

The close-in ($a$ = 0.0864 $\pm$ 0.0006 au) and eccentric ($e$ = 0.32 $\pm$ 0.02) Jupiter-sized planet HD~118203~b was discovered by \citet{2006A&A...446..717D} through 43 RVs acquired with the ELODIE spectrograph. Recently, the TESS satellite revealed that HD~118203~b transits its bright host star \citep{2020AJ....159..243P}. Having a magnitude of V = 8.05 $\pm$ 0.03, HD~118203 is among the 10 brightest stars to have a transiting giant planet. Close-in giant planets such as HD~118203~b may influence the activity levels of their host stars through magnetic star-planet interactions (MSPIs), which may induce stellar variability modulated by the orbital period of the interacting planet. Since the first claims \citep{2003A&A...406..373S,2003ApJ...597.1092S}, dozens of signs of MSPIs have been reported \citep[e.g.][]{2005ApJ...622.1075S,2008ApJ...676..628S,2008A&A...482..691W,2009EM&P..105..373P,2011ApJ...741L..18P,2015ApJ...811L...2M}. However, a key regime remained unexplored: eccentric planetary systems. Eccentric systems hosting close-in giant planets such as HD~118203 are very scarce since at those close distances gas giants suffer a quick circularization process \citep{1981A&A....99..126H}. Previous studies focussed on detecting MSPIs in eccentric systems were unsuccessful, and to date, there are only two possible cases of MSPIs in eccentric systems \citep{2015ApJ...803...49Q,2015ApJ...811L...2M}. However, eccentric systems can give us important insights on MSPIs.

In this work, we analysed the complete ELODIE dataset (43 RVs from the discovery paper and 13 more public RVs) and four TESS sectors of HD~118203 with the primary objective of searching for activity signals potentially induced by its hosted close-in giant planet, and the secondary objective of refining the orbital and physical properties of the planetary system. 

We found evidence of an activity signal within the TESS photometry that matches the orbital period of HD~118203~b (i.e.~$\simeq$6.1 days), which could be the result of magnetic interactions between the planet and its host star. In order to confirm such interactions, the commonly considered criterion consists of independently detecting the rotation period of the star. To do so, we analysed the ELODIE activity indicators and complementary ASAS-SN photometry but found no compelling evidence of an additional rotation-induced activity signal. However, given the evolved nature of the star and the significant orbital eccentricity, we argue that MSPIs is the most likely scenario to explain the observed TESS variability. This argument agrees
with literature estimates of the spectroscopic projected rotational velocity. 

We analysed the persistence and evolution of the signal and found that it appears and disappears in time scales comparable to a planetary orbit. Also, when active, it experiences strong changes in both amplitude and shape. We interpret this behaviour as a possible manifestation of the on/off nature of MSPIs found by \citet{2008ApJ...676..628S}. However, it does not completely rule out the rotation scenario since the star could have a quick spot evolution that could cause those sudden changes \citep[e.g.][]{2019ApJ...871..187N}. We also studied the existence of links between the TESS variability and the orbital motion of HD~118203~b. Unfortunately, the observed variations are complex and cannot be interpreted as a single small spot moving over the stellar surface, which implies that there is a geometric factor that we  cannot consider in our analysis. Overall, we found a moderate degree of correlation between the flux derivative of HD~118203 and the angular velocity of HD~118203~b in two chunks of the TESS photometry but found no correlations in the remaining four chunks.  We also found that the rotation period of the planet is expected to be pseudo-synchronized with the planetary orbital period with a 3.73-day periodicity. Interestingly, such a spin velocity is expected to generate a planetary magnetic moment able to produce strong MSPIs, but in a hypothetical circular scenario in which the planetary rotation and orbital periods were synchronized, the generated magnetic moment could not be enough to produce such interactions \citep{2005ApJ...622.1075S,2008ApJ...676..628S}. This suggests that the unusually high eccentricity of HD~118203~b could be critical for the generation of the potential MSPIs found.

Regarding the secondary objective of the work, we significantly improved the transit ephemeris of this planetary system thanks to the inclusion of two new sectors of TESS data separated 2.5 years from the last observations. As an example, for the year 2028, we compute a propagated uncertainty in the mid-transit time of just 1 minute, in contrast with the 27 minutes uncertainty reported in the current most accurate characterization \citep{2020AJ....159..243P}. Obtaining accurate ephemeris for very bright targets is very valuable for the community since those targets will be most likely scheduled for atmospheric studies.

Similar to previous studies such as \citet{2008A&A...482..691W} and \citet{2009EM&P..105..373P}, this work reports new signs of MSPIs detected photometrically in a planetary system hosting a hot Jupiter, whose confirmation via rotation period determination should be attempted through precise photometric or spectroscopic follow-up observations. In this regard, the PLATO mission \citep{2014ExA....38..249R} will play a major role thanks to the expected long-term and continuous precise photometric monitoring of the sky. Also, state-of-the-art high-resolution spectrographs such as HARPS-N \citep{2012SPIE.8446E..1VC} and CARMENES \citep{2014SPIE.9147E..1FQ} could be very useful to determine the rotation period of this star, either through direct observations of the stellar chromospheric activity or through the Rossiter-McLaughlin effect \citep{1924ApJ....60...15R,1924ApJ....60...22M}.  To date, HD 118203 represents the best evidence that magnetic star-planet interactions can be found in eccentric planetary systems, and it opens the door to future dedicated searches in such systems that will allow us to better understand the interplay between close-in giant planets and their host stars.

\begin{acknowledgements}

We thank the referee for the thorough and constructive comments, which improved the quality of this work. A.C.-G. is funded by the Spanish Ministry of Science through MCIN/AEI/10.13039/501100011033 grant PID2019-107061GB-C61. J.L.-B. is supported by the Spanish Ministry of Science through the Ram\'on y Cajal program with code RYC2021-031640-I (funded through MCIN/AEI/10.13039/501100011033 and the NextGenerationEU/PRTR from the European Union). N.C.S is co-funded by the European Union (ERC, FIERCE, 101052347). Views and opinions expressed are however those of the author(s) only and do not necessarily reflect those of the European Union or the European Research Council. Neither the European Union nor the granting authority can be held responsible for them. This work was supported by FCT - Fundação para a Ciência e a Tecnologia through national funds and by FEDER through COMPETE2020 - Programa Operacional Competitividade e Internacionalização by these grants: UIDB/04434/2020, UIDP/04434/2020, UIDB/04564/2020, UIDP/04564/2020, PTDC/FIS-AST/7002/2020, and POCI-01-0145-FEDER-022217. 
This work is based on spectral data retrieved from the ELODIE archive at Observatoire de Haute-Provence (OHP).

This work made use of \texttt{tpfplotter} by J. Lillo-Box (publicly available in \url{www.github.com/jlillo/tpfplotter}), which also made use of the python packages \texttt{astropy}, \texttt{lightkurve}, \texttt{matplotlib} and \texttt{numpy}. This work has made use of data from the European Space Agency (ESA) mission {\it Gaia} (\url{https://www.cosmos.esa.int/gaia}), processed by the {\it Gaia} Data Processing and Analysis Consortium (DPAC, \url{https://www.cosmos.esa.int/web/gaia/dpac/consortium}). We acknowledge the use of public TESS data from pipelines at the TESS Science Office and at the TESS Science Processing Operations Center. Resources supporting this work were provided by the NASA High-End Computing (HEC) Program through the NASA Advanced Supercomputing (NAS) Division at Ames Research Center for the production of the SPOC data products. This research has made use of the Exoplanet Follow-up Observation Program (ExoFOP; DOI: 10.26134/ExoFOP5) website, which is operated by the California Institute of Technology, under contract with the National Aeronautics and Space Administration under the Exoplanet Exploration Program. This research has made use of the NASA Exoplanet Archive, which is operated by the California Institute of Technology, under contract with the National Aeronautics and Space Administration under the Exoplanet Exploration Program. This research has made use of the SIMBAD database \citep{2000A&AS..143....9W}, operated at CDS, Strasbourg, France. This work has made use of the following software: \texttt{astropy} \citep{2022ApJ...935..167A}, \texttt{matplotlib} \citep{2007CSE.....9...90H}, \texttt{numpy} \citep{2020Natur.585..357H}, \texttt{scipy} \citep{2020NatMe..17..261V}, \texttt{lightkurve} \citep{2018ascl.soft12013L}, and \texttt{corner} \citep{2016JOSS....1...24F}.

\end{acknowledgements}

%
%

\bibliographystyle{aa} 
\bibliography{references} 
\begin{appendix}

\section{Additional tables}


\begin{table*}[]
\renewcommand{\arraystretch}{1.25}
\setlength{\tabcolsep}{28pt}
\caption{Inferred parameters of HD~118203~b obtained from the joint analysis of TESS photometry and ELODIE RVs (Sect.~\ref{sec:joint_fit}).}
\label{tab:bestfit}
\begin{tabular}{lll}
\hline
Parameter                                                       & Priors                          & Posteriors                      \\ \hline
\multicolumn{3}{l}{Orbital parameters}                                                                                              \\ \hline
Orbital period, $P_{\rm orb}$ (days)                            & $\mathcal{U}(6.0, 6.2)$         & $6.1349847 \pm 0.0000020$       \\
Time of mid-transit, $T_{0}$ (JD)                               & $\mathcal{U}(2459663, 2459664)$ & $2459663.58480 \pm 0.00025$     \\
Orbital inclination, $i$ (degrees)                              & $\mathcal{U}(50.0, 90.0)$       & $89.16^{+0.57}_{-0.65}$         \\
Ecc. parametrization, $cos(w) \sqrt{e}$                         & $\mathcal{U}(-1, 1)$            & $-0.499^{+0.014}_{-0.013}$      \\
Ecc. parametrization, $sin(w) \sqrt{e}$                         & $\mathcal{U}(-1, 1)$            & $0.259 \pm 0.029$               \\
Orbital eccentricity, $e$                                       & (derived)                       & $0.316 \pm 0.020$               \\
Argument of periastron, $w$ (degrees)                           & (derived)                       & $152.5 \pm 2.7$                 \\
Time of periastron, $T_{\rm p}$ (JD)                                & (derived)                       & $2453394.251 \pm 0.021$         \\
Transit duration, $T_{14}$ (hours)                              & (derived)                       & $5.631 \pm 0.015$               \\ \hline
Planet parameters                                               &                                 &                                 \\ \hline
RV semi-amplitude, $K$ ($\rm m \, s^{-1}$)                      & $\mathcal{U}(200, 250)$         & $218.0^{+4.0}_{-3.9}$           \\
Planet mass, $M_{p} \, (M_{\rm J})$                             & (derived)                       & $2.282 \pm 0.045$               \\
Scaled planet radius, $R_{p}/ R_{\star}$                        & $\mathcal{U}(0.0, 0.5)$         & $0.05546^{+0.00024}_{-0.00022}$ \\
Planet radius, $R_{p} \, (R_{\rm J})$                           & (derived)                       & $1.076 \pm 0.035$               \\
Planet density, $\rho_{p} \, (\rm g \, cm^{-3})$                & (derived)                       & $2.27 \pm 0.23$                 \\
Transit depth, $\Delta$ (ppt)                                   & (derived)                       & $3.075 \pm 0.025$               \\
Relative orbital separation, $a / R_{\star}$                    & (derived)                       & $9.32 \pm 0.31$                 \\
Orbit semimajor axis, $a$ (au)                                  & (derived)                       & $0.08635 \pm 0.00057$           \\
Planet surface gravity, $g \, (\rm m \, s^{-2})$                & (derived)                       & $48.9 \pm 3.4$                  \\
Impact parameter, $b$                                           & (derived)                       & $0.14 \pm 0.10$                 \\
Incident flux, $F_{\rm inc} \, (F_{\oplus})$                    & (derived)                       & $593.4 \pm 8.2$                 \\
Equilibrium temperature [A=0], $T_{\rm eq} \, (\rm K)$          & (derived)                       & $1360 \pm 23$                   \\ \hline
\multicolumn{3}{l}{Limb-darkening coefficients}                                                                                     \\ \hline
Limb-darkening coefficient, $u_{1}$                             & $\mathcal{U}(0, 1)$             & $0.347^{+0.041}_{-0.042}$       \\
Limb-darkening coefficient, $u_{2}$                             & $\mathcal{U}(0, 1)$             & $0.150^{+0.071}_{-0.069}$       \\ \hline
\multicolumn{3}{l}{GP hyperparameters}                                                                                              \\ \hline
$\eta_{\rm \sigma_{S15}}$ ($\rm e^{-}s^{-1}$)                   & $\mathcal{U}(0, 10^{3})$        & $76.7^{+5.6}_{-4.9}$            \\
$\eta_{\rm \rho_{S15}}$ (days)                                  & $\mathcal{U}(0, 10^{2})$        & $0.138^{+0.010}_{-0.009}$       \\
$\eta_{\rm \sigma_{S16}}$ ($\rm e^{-}s^{-1}$)                   & $\mathcal{U}(0, 10^{3})$        & $48.8^{+4.6}_{-3.9}$            \\
$\eta_{\rm \rho_{S16}}$ (days)                                  & $\mathcal{U}(0, 10^{2})$        & $0.229^{+0.025}_{-0.021}$       \\
$\eta_{\rm \sigma_{S22}}$ ($\rm e^{-}s^{-1}$)                   & $\mathcal{U}(0, 10^{3})$        & $37.1^{+5.7}_{-4.3}$            \\
$\eta_{\rm \rho_{S22}}$ (days)                                  & $\mathcal{U}(0, 10^{2})$        & $0.60^{+0.11}_{-0.09}$          \\
$\eta_{\rm \sigma_{S49}}$ ($\rm e^{-}s^{-1}$)                   & $\mathcal{U}(0, 10^{3})$        & $45.6^{+5.3}_{-4.3}$            \\
$\eta_{\rm \rho_{S49}}$ (days)                                  & $\mathcal{U}(0, 10^{2})$        & $0.310^{+0.048}_{-0.039}$       \\ \hline
\multicolumn{3}{l}{Instrument-dependent parameters}                                                                                 \\ \hline
TESS LC level S15, $F_{\rm 0,S15}$ ($\rm e^{-}s^{-1}$)          & $\mathcal{U}(-10^{3}, 10^{3})$  & $11.3^{+9.1}_{-9.3}$            \\
TESS LC level S16, $F_{\rm 0,S16}$ ($\rm e^{-}s^{-1}$)          & $\mathcal{U}(-10^{3}, 10^{3})$  & $8.1^{+7.6}_{-7.8}$             \\
TESS LC level S22, $F_{\rm 0,S22}$ ($\rm e^{-}s^{-1}$)          & $\mathcal{U}(-10^{3}, 10^{3})$  & $6.7^{+8.5}_{-8.3}$             \\
TESS LC level S49, $F_{\rm 0,S49}$ ($\rm e^{-}s^{-1}$)          & $\mathcal{U}(-10^{3}, 10^{3})$  & $5.4^{+8.6}_{-8.5}$             \\
TESS LC jitter S15, $\sigma_{\rm TESS,S15}$ ($\rm e^{-}s^{-1}$) & $\mathcal{U}(0, 10^{2})$        & $30.72 \pm 0.60$                \\
TESS LC jitter S16, $\sigma_{\rm TESS,S16}$ ($\rm e^{-}s^{-1}$) & $\mathcal{U}(0, 10^{2})$        & $31.79 \pm 0.63$                \\
TESS LC jitter S22, $\sigma_{\rm TESS,S22}$ ($\rm e^{-}s^{-1}$) & $\mathcal{U}(0, 10^{2})$        & $27.83 \pm 0.59$                \\
TESS LC jitter S49, $\sigma_{\rm TESS,S49}$ ($\rm e^{-}s^{-1}$) & $\mathcal{U}(0, 10^{2})$        & $36.81 \pm 0.57$                \\
ELODIE RV jitter, $\sigma_{\rm ELODIE}$ ($\rm m \, s^{-1}$)     & $\mathcal{U}(0, 10^{2})$        & $14.5^{+2.8}_{-2.5}$            \\ \hline
\multicolumn{3}{l}{RV linear trend}                                                                                                 \\ \hline
Systemic velocity, $v_{\rm sys}$ ($\rm m \, s^{-1}$)            & $\mathcal{U}(-30000, -29000)$   & $-29364.3 \pm 4.5$              \\
Slope, $\gamma$ ($\rm m \, s^{-1} \, day^{-1} $)                & $\mathcal{U}(-1, 1)$            & $0.139 \pm 0.011$               \\ \hline
\end{tabular}
\end{table*}

\section{Additional figures}

\begin{figure*}
    \includegraphics[width=\textwidth]{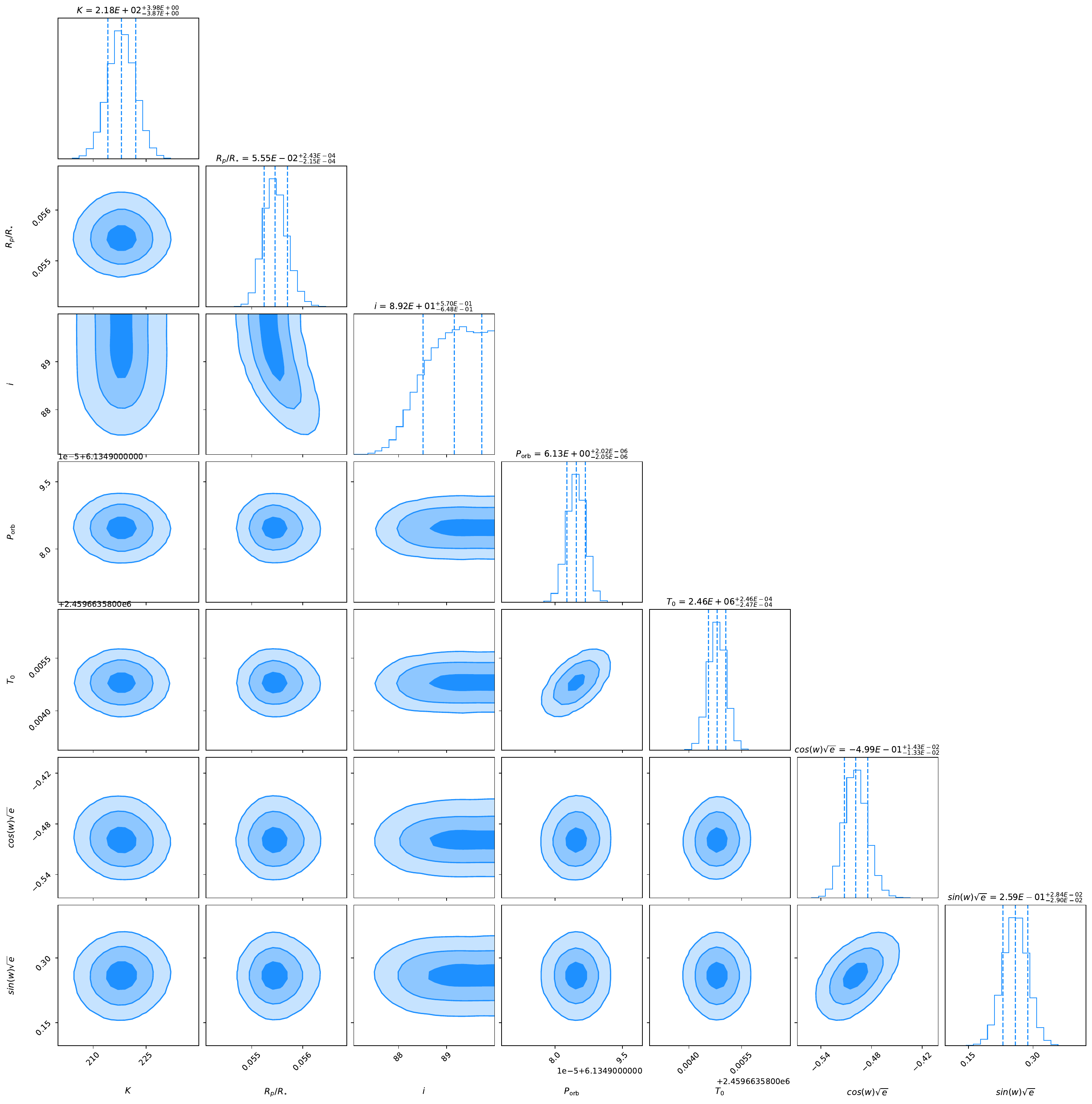}
    \caption{Corner plot of the main parameters describing the HD~118203 planetary system obtained from the joint TESS light curve and ELODIE RV analysis.}
    \label{fig:corner_plot}
\end{figure*}

\begin{figure*}
    \includegraphics[width = 0.5\textwidth]{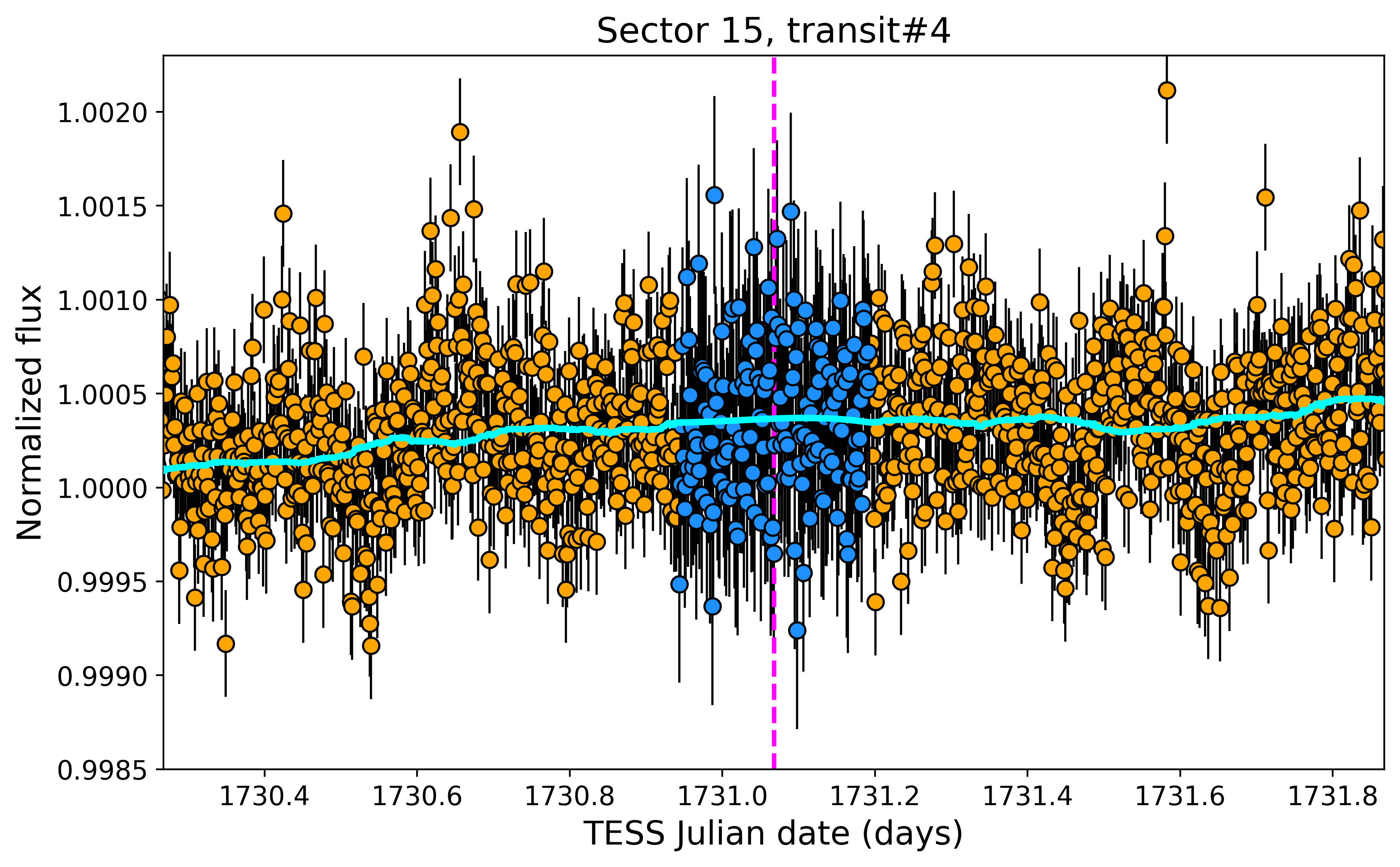}
    \includegraphics[width = 0.5\textwidth]{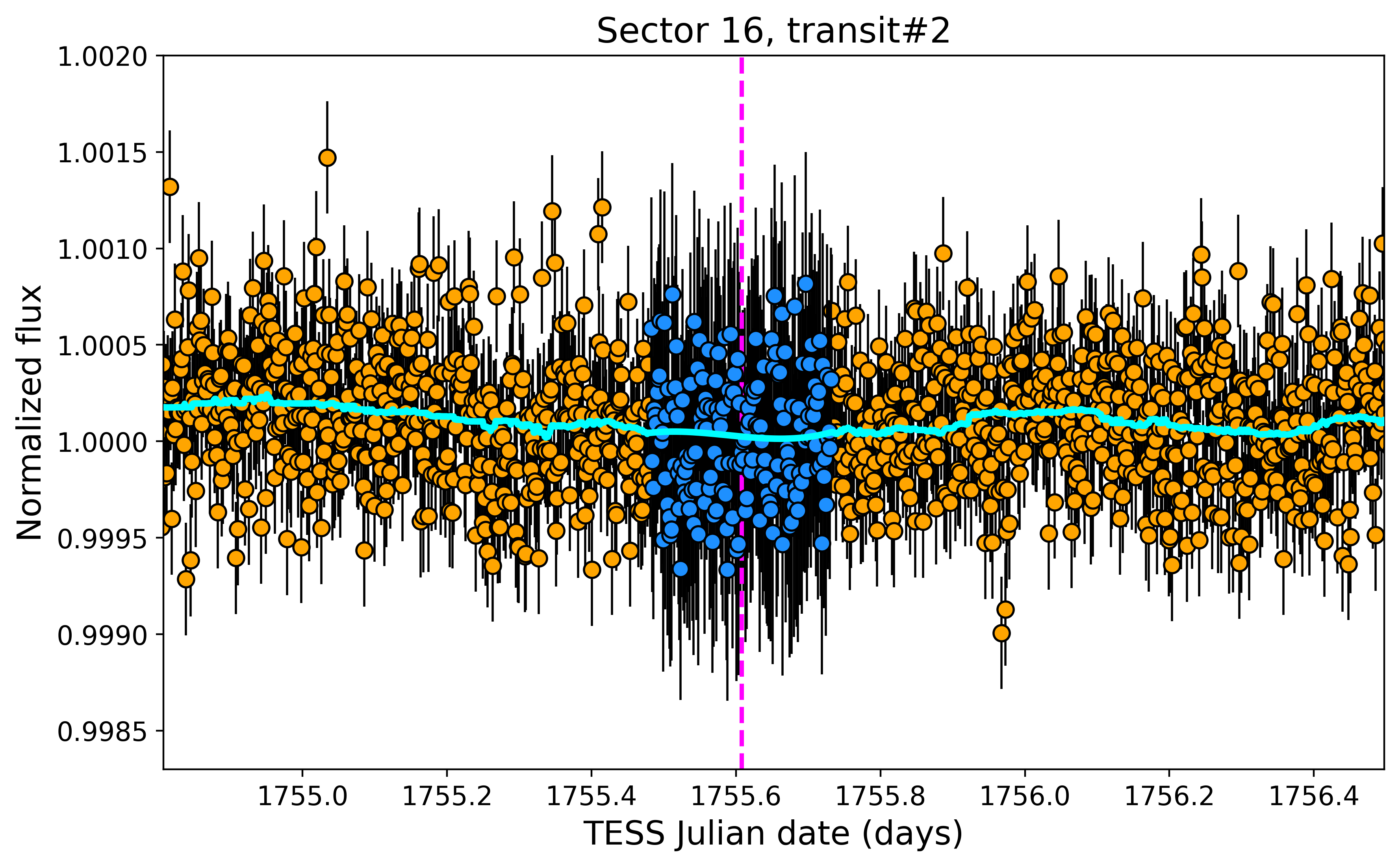}
    \caption{Examples of the mock dataset filling the TESS transit gaps. Orange data represent the TESS PDCSAP photometry. Blue data are the simulated 2-min cadence photometry filling the gaps of the masked transits. The cyan curve represents the cubic spline interpolation model. The vertical magenta dashed line indicates the mid-transit time of the masked transit. }
    \label{fig:mock_data_example}
\end{figure*}

\begin{figure*}
    \includegraphics[width = 0.5\textwidth]{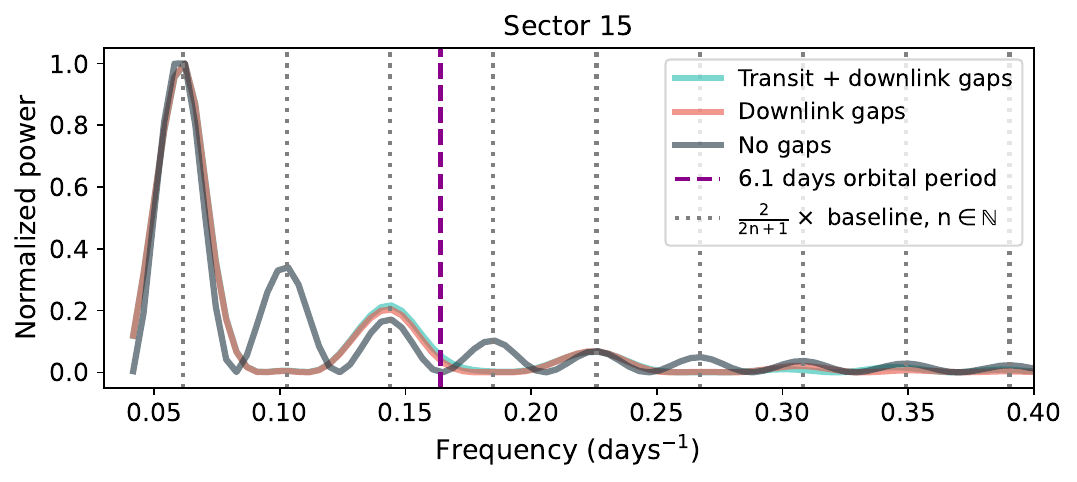}
    \includegraphics[width = 0.5\textwidth]{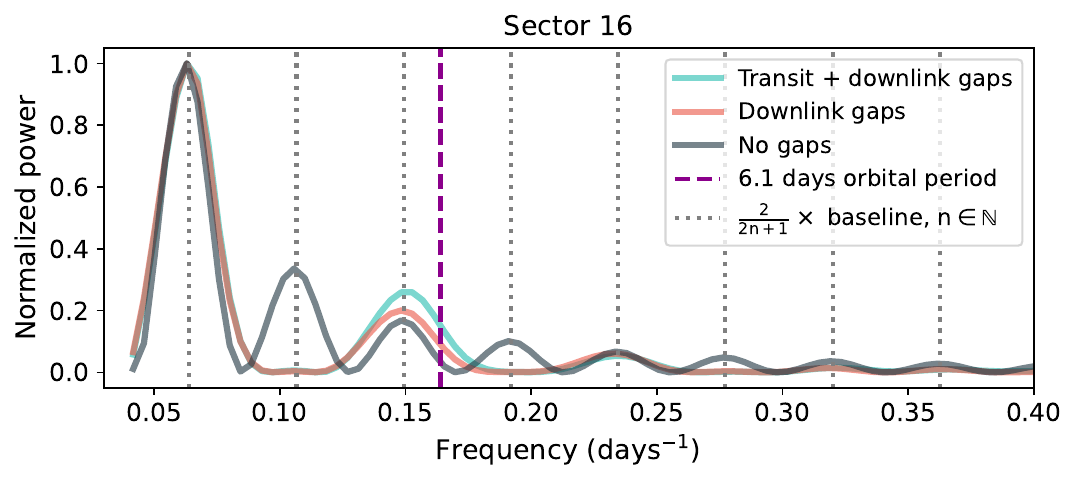}
    \includegraphics[width = 0.5\textwidth]{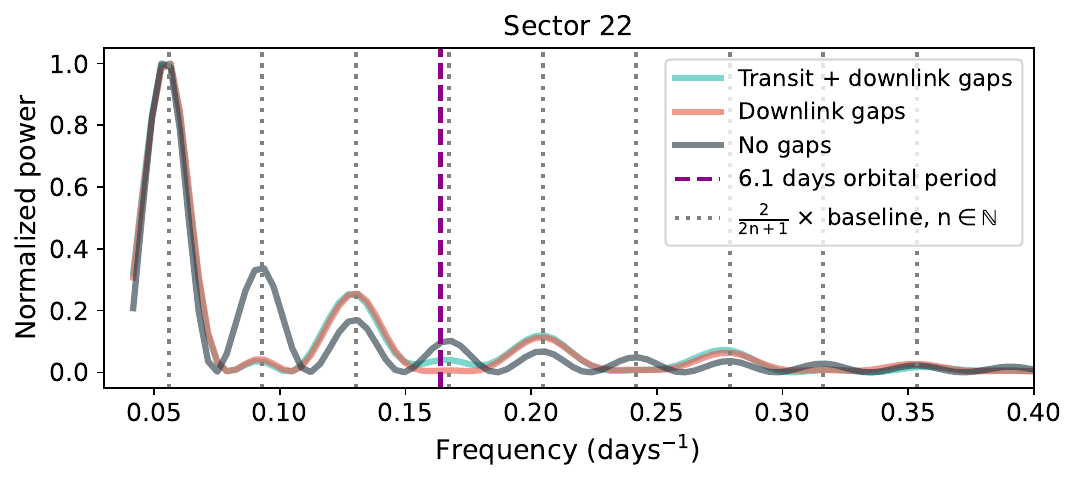}
    \includegraphics[width = 0.5\textwidth]{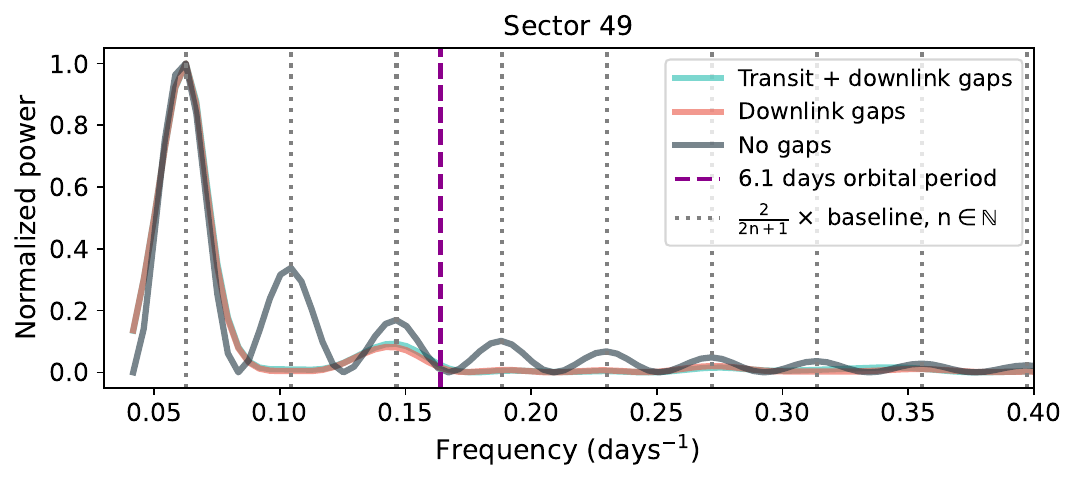}
    \caption{\texttt{GLS} periodograms of the TESS window functions with transit and downlink gaps (blue), downlink gaps alone (red), and no gaps (grey). In all cases, the maximum power periods correspond to two-thirds of the total sector baselines, which indicates that those periodicities are not related to the gaps.}
    \label{fig:wf_comparison}
\end{figure*}


\end{appendix}

\end{document}